\documentclass[journal]{IEEEtran}
\IEEEoverridecommandlockouts
\usepackage{cite}
\usepackage{amsmath,amssymb,amsfonts,amsthm}
\usepackage{algorithmic}
\usepackage{graphicx}
\usepackage{textcomp}
\usepackage{xcolor}
\usepackage{float}
\usepackage{bm}
\usepackage{bbold}
\usepackage{tcolorbox}
\usepackage{diagbox}
\usepackage{tabularx}
\usepackage{multirow}
\usepackage{multicol}
\usepackage{subcaption}
\usepackage{lettrine}
\usepackage[normalem]{ulem}
\newcommand{\stkout}[1]{\ifmmode\text{\sout{\ensuremath{#1}}}\else\sout{#1}\fi}
\usepackage{enumitem}

\usepackage[nolist,nohyperlinks]{acronym}

\newtheorem{proposition}{Proposition}
\newtheorem{definition}{Definition}

\usepackage{threeparttable}
\usepackage[linesnumbered,ruled,vlined]{algorithm2e}
\usepackage[absolute,overlay]{textpos}
\usepackage{tikz}
\usetikzlibrary{shapes.geometric, arrows.meta, positioning, fit, calc}
\graphicspath{ {./figures/}}
\def\BibTeX{{\rm B\kern-.05em{\sc i\kern-.025em b}\kern-.08em
    T\kern-.1667em\lower.7ex\hbox{E}\kern-.125emX}}

\tikzstyle{startstop} = [ellipse, minimum width=1cm, minimum height=0.5cm,text width=4.5em, text width=4.5em, text centered, draw=black]
\tikzstyle{decision} = [diamond, draw, inner sep=0.45em, fill=white, text width=14em, text centered, node distance=2cm]
\tikzstyle{block} = [rectangle, draw, fill=white, text width=20em, inner sep=1em, rounded corners, minimum height=4em]
\tikzstyle{blockprocess} = [rectangle, draw, fill=white, text width=12em, text centered, rounded corners, minimum height=4em]
\tikzstyle{line} = [draw, -{Latex[length=2mm]}]
\tikzstyle{arrow} = [thick,->,>=stealth]
\tikzstyle{output} = [trapezium, trapezium left angle=70, trapezium right angle=110, minimum width=1cm, minimum height=0.5cm, text centered, draw=black]

\newcommand{\wrt}{\textit{w.r.t.~}}

\newcommand{\ie}{\textit{i.e.~}}

\newcommand{\eg}{\textit{e.g.~}}
\newcommand{\egn}{\textit{e.g.}}
\newcommand{\fig}[1]{Fig. \ref{#1}}
\newcommand{\mbs}{\rm Mb/s}
\newcommand{\m}{\rm m}

\definecolor{lightgray}{gray}{0.35}

\newcommand{\param}{{\mathbf{C}}}
\DeclareMathSymbol{\mathbbE}{\mathord}{AMSb}{"45}

\DeclareMathOperator*{\argmax}{arg\,max}

\begin{document}

\title{Impact of Sociality Regimes on Quality of Service and Energy Efficiency in Cell-Free MIMO Networks} 

\author{Ala Eddine Nouali, Mohamed Sana, and Jean-Paul Jamont}

\author{\IEEEauthorblockN{Ala Eddine Nouali\thanks{An earlier version of this paper was presented in part at the IEEE PIMRC, Valencia, Spain, September 2024 \cite{nouali2024sociality}.}\thanks{Ala Eddine Nouali is with CEA-Leti, Univ. Grenoble Alpes, F-38000 Grenoble, France (e-mail: ala-eddine.nouali@cea.fr).}, Mohamed Sana\thanks{Mohamed Sana was with CEA-Leti when this manuscript was submitted. He is now with Huawei Technologies, Fourier Research Center, 20 quai du Point du Jour, Boulogne Billancourt, France (e-mail: mohamed.sana@huawei.com).}, Jean-Paul Jamont\thanks{
Jean-Paul Jamont is with LCIS, Univ. Grenoble Alpes, Grenoble INP, Valence, France (e-mail: jean-paul.jamont@univ-grenoble-alpes.fr).}} \thanks{This work was funded by the French government under the France 2030 ANR program “PEPR Networks of the Future” (ref. ANR-22-PEFT-0003).}}

\begin{acronym}
\acro{AN}{Artificial Noise}
\acro{OPALS}{Out-Phased Array Linearized Signaling}
\acro{AP}{Access Point}
\acro{UE}{User Equipment}
\acro{CPU}{central processing unit}
\acro{DL}{downlink}
\acro{UL}{uplink}
\acro{RU}{Radio Unit}
\acro{CU}{Central Unit}
\acro{DU}{Distributed Unit}
\acro{ULA}{Uniform Linear Array}
\acro{AoA}{angle of arrival}
\acro{RT}{ray-tracing}
\acro{CF-MIMO}{cell-free multiple-input-multiple-output}

\acro{SU}{Single-User}
\acro{MU}{Multi-User}
\acro{MIMO}{multiple-input-multiple-output}
\acro{TDD}{Time Division Duplexing}
\acro{OFDM}{Orthogonal Frequency Division Multiplexing}
\acro{CSI}{Channel State Information}
\acro{NMSE}{Normalized Mean Square Error}

\acro{PAPR}{Peak-to-average Power Ration}
\acro{SNR}{Signal to Noise Ratio}
\acro{SINR}{Signal-to-Interference-Noise Ratio}
\acro{DM}{Dynamic Modulation}
\acro{AN}{Artificial Noise}
\acro{EF}{Energy Efficiency}

\acro{ZF}{Zero forcing}
\acro{SVD}{Singular Value decomposition}
\acro{SINR}{Signal-to-Interference-plus-Noise Ratio}
\acro{QoS}{quality of service}
\acro{EE}{Energy Efficiency}
\acro{SE}{Spectral Efficiency}
\acro{LoS}{line-of-sight}
\acro{NLoS}{non-line-of-sight}
\acro{PE}{Power Efficiency}
\acro{CDF}{Cumulative Distribution Function}
\acro{DDM}{Dynamic Directional Modulation}

\acro{UC}{user-centric}
\acro{UCC}{user-centric clustering}
\acro{DA}{Deferred Acceptance}
\acro{EA}{Early Acceptance}
\end{acronym}

\maketitle
\begin{abstract}
The cell-free architecture represents a significant advancement in network design, where each \ac{UE} is served by a group of distributed \acp{AP}, aimed at delivering uniformly high data rates to \acp{UE} across all locations. 
To ensure network scalability, \ac{UCC} has emerged as a practical approach, wherein only a selected subset of preferred \acp{AP} jointly serves each \ac{UE}. Forming the optimal cluster of \acp{AP} for each \ac{UE} is a challenging task, particularly when limited fronthaul and processing capabilities of both \acp{AP} and \acp{UE} are considered. This challenge is exacerbated by the need for dynamic adjustments to enhance energy efficiency while meeting \ac{QoS} requirements, which introduces conflicting interests between these entities. In this paper, we investigate the sociality regime of \acp{UE} and their clusters of \acp{AP}, characterizing their intra-team cooperation as either selfish, egalitarian, or altruistic. These sociality regimes are crucial in achieving a balance between \ac{QoS} and energy efficiency per \ac{UE}. We address this problem by modeling it as a many-to-many social matching game with externalities, where connections can be established based on the sociality regime of both teams. To solve this, we introduce two novel algorithms based on \ac{DA} and \ac{EA} matching games. Numerical results show the significant impact of the sociality regime adopted by \acp{UE} and their clusters of APs on \ac{UE}'s \ac{QoS} satisfaction and energy efficiency, with the egalitarian regime adopted by both entities proving the best performance trade-off. 
\end{abstract}

\section{Introduction}\label{sec:introduction}

\lettrine[lines=2, lhang=0.1, loversize=0.2]{$\mathbf{T}$}{o} meet the surging demand for higher data rates, both research and industry are increasingly turning their attention to 6G, the next-generation mobile communication technology. 6G promises significant enhancements in capacity, data rate, and latency, and is seen as the natural evolution of current 5G networks. As experts explore this evolution, discussions are intensifying around innovative new architectures and technologies that could shape the future of telecommunications \cite{On_the_road_to_6G_Visions_requirements_key_technologies_and_testbeds}. One of the most promising developments in this area is the \ac{CF-MIMO} network architecture. Unlike traditional cellular networks, where users are served within cells, \ac{CF-MIMO} networks operate through a distributed network of \acp{AP} \cite{Cell_free_massive_MIMO_versus_small_cells}. This distributed approach not only enhances beamforming capabilities and energy efficiency but also aims to provide a uniform high \ac{QoS} to \acp{UE} across the network, regardless of their locations. To guarantee the scalability of the network, \ac{UCC} emerges as an effective concept, where each UE connects to a subset of nearby preferred \acp{AP} \cite{Cell_free_massive_MIMO_User_centric_approach, Foundations_of_user_centric_cell_free_massive_MIMO}. Thus, this approach reduces the fronthaul and backhaul requirements compared to the canonical setup, where all \acp{AP} connect to all UEs. However, despite the potential benefits, the UC approach faces several challenges, particularly in terms of processing capabilities and fronthaul capacities of both \acp{AP} and \acp{UE} \cite{Power_efficient_transmission_for_user_centric_networks_with_limited_fronthaul_capacity_and_computation_resource}. The distributed nature of UC networks demands sophisticated hardware from \acp{AP} and UEs, including advanced signal processors and clock circuits, to handle the complex task of managing multiple exchanged signals effectively. Moreover, the UC approach does not inherently ensure that network resources such as signal processing power, fronthaul/backhaul signaling capacity, and overall energy consumption scale independently with the number of UEs. As a result, when the number of \acp{UE} increases significantly, the network may struggle to maintain service quality, making the UC approach less viable for large-scale deployments. To address these limitations, one potential approach is to restrict the number of \acp{UE} each AP can serve and the number of \acp{AP} to which each UE can connect. This strategy could help alleviate the processing demands on \acp{AP} and \acp{UE} and improve the scalability of \ac{UCC}, making it more practical for ultra-dense networks.

Considering the dynamic nature of UEs traffic requests and their varying QoS requirements, finding the optimal clusters of \acp{AP} for each UE becomes increasingly complex. Moreover, \ac{CF-MIMO} network can be seen as a multi-agent system, where both APs and UEs act as intelligent agents making decisions under limited radio resources. These agents exhibit diverse objectives: \acp{AP} may need to balance their energy efficiency, while UEs may either pursue their own QoS maximization or make trade-offs to support other \acp{UE}. These interaction dynamics arising from this competition and cooperation are highly relevant to real-world scenarios, particularly under heterogeneous QoS demands, requiring a strategic balance to satisfy the needs of both network entities.

\subsection{Background and Related Works}
In the context of \ac{CF-MIMO} networks, \ac{UCC} has emerged as an effective approach to ensure network scalability while optimizing key parameters such as perceived QoS and network energy efficiency. In \cite{Whale_Swarm_Reinforcement_Learning_Based_Dynamic_Cooperation_Clustering_Method_for_Cell_Free_Massive_MIMO_Systems, User_Centric_Clustering_in_Cell_Free_MIMO_Networks_using_Deep_Reinforcement_Learning}, the cluster formation is performed to maximize the total network throughput.
In \cite{Deep_reinforcement_learning_for_dynamic_access_point_activation_in_cell_free_MIMO_networks, User_Centric_Clustering_in_Cell_Free_MIMO_Networks_using_Deep_Reinforcement_Learning}, the authors propose to minimize the number of active \acp{AP} to save energy, while maintaining a good QoS for UEs. \cite{Joint_AP_On_Off_and_User_Centric_Clustering_for_Energy_Efficient_Cell_Free_Massive_MIMO_Systems} deactivates \acp{AP} to achieve the trade-off between the minimum of perceived data rate by \acp{UE} and network energy efficiency. Authors of
\cite{Joint_power_allocation_and_load_balancing_optimization_for_energy_efficient_cell_free_massive_MIMO_networks} propose sleep mode mechanisms and power allocation strategies to minimize network power consumption while guaranteeing a minimum QoS for each UE. The approach in \cite{Matched_decision_AP_selection_for_user_centric_cell-free_massive_MIMO_networks} forms intermediate clusters based on large-scale fading. Then, \acp{UE} expand their clusters of \acp{AP} by connecting to more unsaturated APs. In the final step, these clusters are fine-tuned to form optimized clusters. This fine-tuning strategy aims to reduce the number of \acp{UE} per AP while maintaining perceived data rate or improving energy efficiency.

However, while previous work limits the number of \acp{UE} that an \ac{AP} can serve, and other studies restrict the number of \acp{AP} that can serve a single UE, no existing work simultaneously restricts both \acp{AP} and \acp{UE} to take into account the processing and fronthaul capacities of those network components. By considering both AP and UE capacity constraints, future research could optimize network performance and resource allocation. Such an approach would not only improve network performance but also enhance its scalability. Addressing this dual restriction is therefore critical for the development of more efficient and sustainable network designs.
Furthermore, these works do not specifically consider the requirements of UEs, leading to non-optimal resource allocation. In fact, each user has specific needs, with some \acp{UE} requiring higher data rates and others needing less, resulting in heterogeneous QoS requirements. Therefore, radio resources should be allocated based on these specific needs to avoid over-allocating resources to \acp{UE} that require less and under-allocating to those that need more.

Given the constraints of limited radio resources and QoS requirements of UEs, a diverse array of strategies has been explored to optimize the allocation of radio resources. Among these, the many-to-many matching game has emerged as a particularly robust approach \cite{Matching_theory_for_future_wireless_networks_Fundamentals_and_applications}. Based on the principles of matching theory, this approach has captured the interest of researchers due to its ability to enable distributed user association with low computational complexity and fast convergence time, making it a highly suitable choice for scalable network implementations.
Recent advancements in applying matching theory to resource allocation in wireless networks have increasingly emphasized the role of sociality regimes. In these models, participants not only pursue their individual objectives but also take into account the welfare of others \cite{Dynamic_clustering_and_user_association_in_wireless_small_cell_networks_with_social_considerations, On_social_aware_content_caching_for_D2D-enabled_cellular_networks_with_matching_theory}, with some exhibiting behaviors that range from mutual concern to full altruism \cite{altruistic_players2023}. This shift towards considering social regime in wireless network optimization reflects a growing understanding that real-world interactions often extend beyond pure self-interest, leading to more robust and equitable resource distribution strategies.

\subsection{Our Contributions}

The main contributions of this paper can be summarized as follows:
\subsubsection{Joint QoS Satisfaction and Energy Efficiency per Cluster Maximization} 
We address the \ac{UCC} problem with a dual objective: jointly maximizing the QoS satisfaction for each UE and the energy efficiency of its cluster of APs. Unlike existing works, our approach explicitly accounts for the fronthaul capacities and processing capabilities of network entities and considers the diverse QoS requirements of UEs. This approach ensures that the clustering process not only meets energy efficiency targets but also maintains a good service quality experienced by users, allowing for a more balanced and effective clustering strategy.

\subsubsection{User association under sociality regimes}
We investigate the impact of cooperation levels within intra-UE and intra-AP clusters on both per-UE QoS satisfaction and the energy efficiency of each cluster. Building on the framework proposed in \cite{sociality_regime}, we develop two mathematical models to represent the social behavior of UEs and APs, respectively. Each model incorporates a sociality parameter—ranging from 0 (fully altruistic) to 1 (fully selfish), which quantifies the degree of cooperation among \acp{UE} and \acp{AP}. While we explore the effects of varying these parameters to capture a spectrum of behavioral dynamics, we focus our in-depth analysis on three basic sociality regimes: selfish, egalitarian, and altruistic. These regimes reflect key cooperation-competition scenarios that are highly relevant for practical use-cases, and provide a structured, interpretable framework to analyze how user association strategies affect network performance under different social behaviors.

\subsubsection{Matching Game Based User Association Scheme}
We first cast the problem into a many-to-many \emph{social} matching game \cite{Matching_theory_for_future_wireless_networks_Fundamentals_and_applications}, during which a UE or a cluster of \acp{AP} may accept or reject a matching operation depending on its sociality regime. More specifically, following \cite{sociality_regime}, we investigate nine combinations of three different sociality regimes adopted by \acp{UE} and clusters of \acp{AP}, in which \acp{UE} and clusters of \acp{AP} are either selfish, egalitarian, or altruistic. Hence, we propose two novel algorithms based on \ac{DA} and \ac{EA} matching games. These two algorithms have demonstrated their effectiveness in addressing the \ac{UCC} problem in our previous works 
\cite{Early_Acceptance_Matching_Game_for_User_Centric_Clustering_in_Scalable_Cell_free_MIMO_Networks, nouali2024sociality}. In the DA game, \acp{AP} buffer UE association requests based on their preferences, establishing associations at the end of the game. In contrast, in the EA game, the AP decides immediately to accept or reject a UE upon receipt of its request, which accelerate the matching process.

The rest of this paper is organized as follows. Section \ref{sec:system_model} describes the system model including the channel model and the network energy consumption model. Section \ref{sec:sociality_regimes} introduces the sociality regime in QoS satisfaction per UE and energy efficiency of its cluster of APs, followed by the formulation of a non-convex problem for \ac{UCC}. We detail our matching game based algorithms in section \ref{sec:M2M}. We evaluate the performance of our proposed solutions via simulations in Section \ref{sec:results}. Finally, Section \ref{sec:conclusion} concludes the paper and gives some perspectives on our future work.

\section{System model}\label{sec:system_model}

\subsection{Channel model}
\begin{figure}[!t]
    \centering
    \includegraphics[width=\columnwidth]{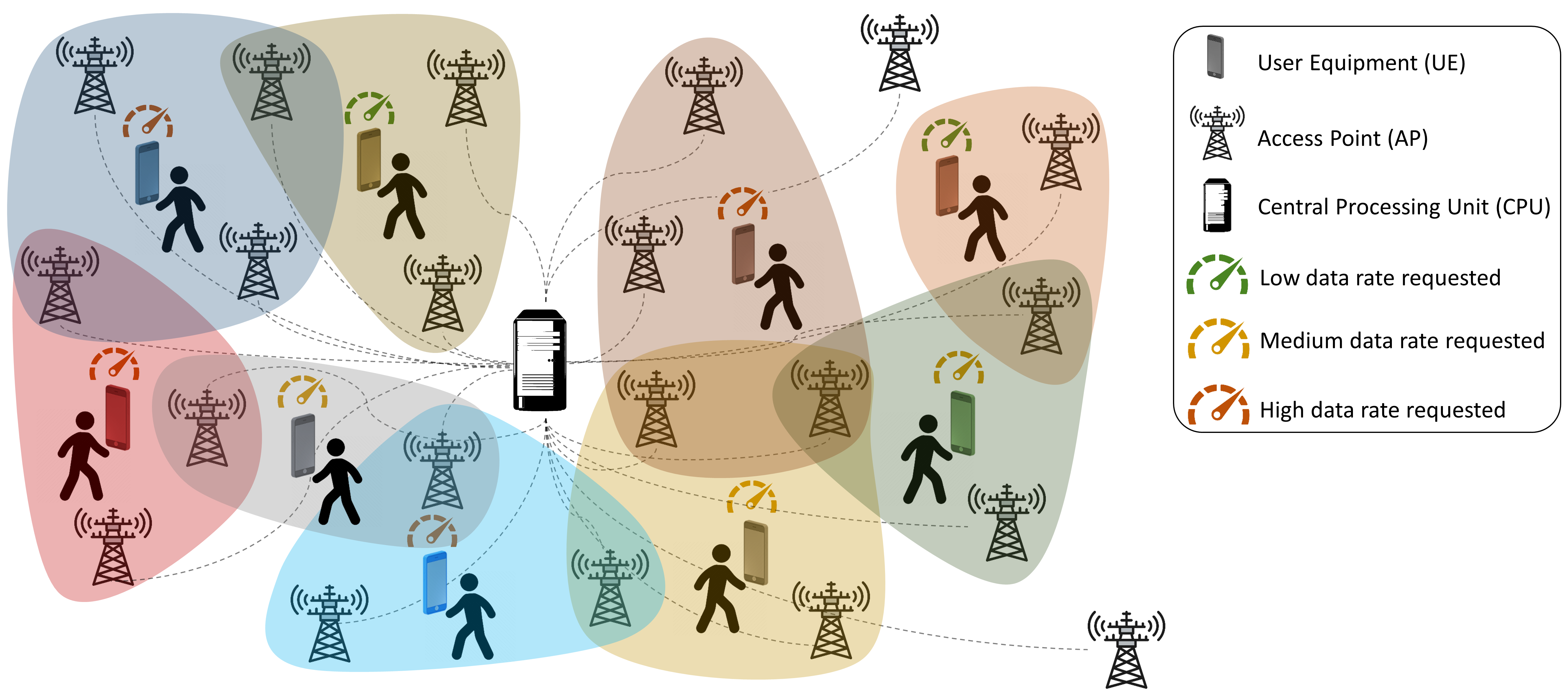}
    \caption{A \ac{UCC} operation in a \ac{CF-MIMO} network with $M$ geographically distributed \acp{AP} connected to a CPU via fronthaul links. The \acp{AP} cooperate to jointly serve $K$ geographically distributed \acp{UE}. Each \ac{UE} is served by a cluster of nearby preferred \acp{AP} depending on its \ac{QoS} requirements.}
    \label{fig:user-centric clustering}
\end{figure}

We consider a downlink \ac{CF-MIMO} network consisting of $M$ geographically distributed APs, where each AP $m$ is equipped with a uniform linear antenna array (ULA) composed of $N$ antenna elements. In the considered system model, a central processing unit (CPU) connects the \acp{AP} via fronthaul links to serve $K$ geographically distributed single-antenna \acp{UE} with different QoS requirements as illustrated in \fig{fig:user-centric clustering}. We denote by $\mathcal{M} = \{1,\dots,M\}$ and $\mathcal{K} = \{1,\dots,K\}$ the set of \acp{AP} and UEs, respectively. We consider spatially correlated Rician fading channels as in \cite{Massive_MIMO_with_spatially_correlated_Rician_fading_channels}, where the channel vector $\mathbf{h}_{k,m} \in \mathcal{C}^{N \times 1}$ between UE $k$ and AP $m$ is modeled as follows:
\begin{equation}\label{eq:channel-model}
    \mathbf{h}_{k,m} \sim \mathcal{N}_{\mathbb{C}}\left(\sqrt{\kappa_{k,m}}\bar{\mathbf{h}}_{k,m}, \mathbf{R}_{k,m}\right) \times \sqrt{\frac{g_{k,m}}{\kappa_{k,m} + 1} }.
\end{equation}
In Eq. \eqref{eq:channel-model}, $\kappa_{k,m}$ denotes the Rician factor and $g_{k,m}$ is the average path-loss, which is function of the distance $d_{k,m}$ between AP $m$ and UE $k$. Also, $\bar{\mathbf{h}}_{k,m} = \mathbb{E}[\mathbf{h}_{k,m}]\in \mathcal{C}^{N \times 1}$  describes the array response vector of the \ac{LoS} component for an azimuth \ac{AoA} $\varphi_{k,m}$ from AP $m$ to UE $k$, \wrt the broadside direction of the AP $m$ antenna array. Its expression reads as:
\begin{equation}\label{eq:LoS_component}
    \bar{\mathbf{h}}_{k,m} = \left[1, e^{j 2\pi\frac{\Delta\sin{(\varphi_{k,m})}}{\lambda}}, \dots, e^{j 2\pi\frac{\Delta(N-1)\sin{(\varphi_{k,m})}}{\lambda}}\right]^{T}
\end{equation}
Here, $\Delta$ denotes the antenna spacing, $\lambda$ is the wavelength of the carrier frequency $f$ and $[.]^{T}$ is the transpose operator. Also, $\mathbf{R}_{k,m} \in \mathcal{C}^{N \times N}$ denotes the positive semi-definite covariance matrix describing the spatial correlation of the \ac{NLoS} components between AP $m$ and UE $k$ \cite{Massive_MIMO_with_spatially_correlated_Rician_fading_channels}.

\subsection{User Association and Data Rate}

\begin{figure}[!t]
    \centering
    \resizebox{\columnwidth}{!}{
    \begin{tikzpicture}[
          cline/.style={help lines,black},
          carc/.style={>=Triangle,cline,<->},
          cbox/.style={rectangle, inner sep=2mm, line width=1pt, minimum width=2cm, minimum height=0.6cm, text centered}]
        ]
        \path[inner sep=0]
          node[cbox, fill=gray!30] (nA) {\footnotesize Association Phase} 
          node[cbox, fill=red!30, right=0pt of nA] (nB) {\footnotesize Data Transmission Phase}
          node[inner sep=2mm, right=0pt of nB] (nC) {\ldots}
          node[draw,fit={(nA) (nC)}] (frame) {};
          
        \draw[cline] ([yshift=20pt]frame.north west) -- ([yshift=-20pt]frame.south west)
          (nA.north east|-frame.north) -- ([yshift=-20pt]{nA.south east|-frame.south}) node[below=2pt]{} 
          ([yshift=20pt]{nB.north east|-frame.north}) -- ([yshift=-20pt]{nB.south east|-frame.south}) node[below=2pt]{};
          
        \draw[carc] ([yshift=-10pt]frame.south west) -- node[fill=white] {\footnotesize $T_I$} ([yshift=-10pt]{nA.south east|-frame.south}); 
        \draw[carc] ([yshift=10pt]frame.north west) -- node[fill=white] {\footnotesize Association block $(T_I + T_F)$} ([yshift=10pt]{nB.north east|-frame.north});  
    \end{tikzpicture}
}
    \caption{User association frame structure.}
    \label{fig:frame}
\end{figure}
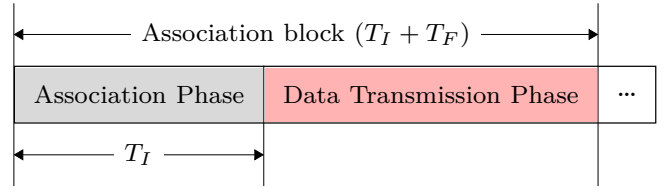

In our system model, a UE is served by a group of APs. As in \cite{Distributed_user_association_in_B5G_networks_using_early_acceptance_matching_game}, we perform user association per time duration called \emph{association block}. An association block can span one or several time slots depending on the coherence time of the channels. An association block comprises an association phase of duration $T_I$ and a data transmission phase of duration $T_F$ as illustrated in Fig. \ref{fig:frame}.
During an association phase, \acp{UE} and \acp{AP} perform beam training and alignment mechanisms, defining for each UE its connected cluster of \acp{AP} and configuring the appropriate beams (channel precoding) \wrt serving \acp{AP} for the data transmission phase. The association phase is further divided into time slots during which \acp{UE} can request association with their surrounding \acp{AP}. A UE request may get rejected by an \ac{AP} due to the unavailability of the radio resources (\eg beams) or the adopted behavior. In this case, it may request an association with other APs. Thus, the duration of the association phase ($T_I$) may vary from an association block to another depending, \egn, on the \ac{CSI} and the \acp{UE} requirements and behaviors. An iterative procedure follows, until a cluster of \acp{AP} is identified for all \acp{UE}. This work focuses on forming such clusters per association block\footnote{In our work, we do not consider the dependence between consecutive association blocks. Accordingly, our problem formulation does not consider a time index.}.

Accordingly, given an association block, we denote by $\mathbf{C} = \{c_{k,m},\;\forall (k,m) \in \mathcal{K}\times \mathcal{M}\}$ the clustering matrix between the $K$ \acp{UE} and $M$ APs, which fully determines the user association in the corresponding association block. 
The term $c_{k,m}$ in $\mathbf{C}$ indicates whether UE $k$ is connected to AP $m$, in which case $c_{k,m} =1$; otherwise $c_{k,m} = 0$. 

In the described scenario, the downlink communication links interfere with each other making the signal-to-interference-plus-noise ratio (SINR) perceived by \acp{UE} highly dependent on the clustering operation. Hence, the SINR of UE $k$ given $\mathbf{C}$ is evaluated as:
\begin{equation} \label{eq:SINR}
    \mathrm{SINR}_k(\param) = \frac{S_k(\param)}{I_k(\param)+\sigma_n^2}.
\end{equation}
Here, $S_k(\param)$ is the desired signal power received by UE $k$, which reads as:
\begin{equation} \label{eq:received_power}
    S_k(\param) = \left|\displaystyle\sum_{m=1}^M \mathbf{h}_{k, m}^H \mathbf{w}_{k, m} c_{k, m}\right|^2,
\end{equation}

Also, $I_k(\param)$ is the perceived interference power given as:
\begin{equation}\label{eq:interference}
    I_k(\param) = \displaystyle\sum_{\substack{j=1 \\ j \neq k}}^K\left|\displaystyle\sum_{m=1}^M \mathbf{h}_{k, m}^H \mathbf{w}_{j, m} c_{j, m}\right|^2.
\end{equation}
In Eq. \eqref{eq:SINR}, $\sigma_{n}^2$ denotes the receiver noise power in a bandwidth $B$ and $\mathbf{w}_{k,m} \in \mathcal{C}^{N \times 1}$ represents the normalized precoding vector assigned by AP $m$ to UE $k$ computed as

\begin{equation}\label{eq:normalized_precoding}
 \mathbf{w}_{k, m} = \sqrt{P_{k, m}^{\rm tx}} \frac{\mathbf{v}_{k,m}}{\sqrt{\mathbb{E}\left\{||\mathbf{v}_{k,m}||^{2}\right\}}},    
\end{equation}
where $P_{k,m}^{\rm tx}$ is the transmit power that AP $m$ allocates to UE $k$,  $\mathbf{v}_{k, m} / \sqrt{\mathbb{E}\left\{||\mathbf{v}_{k,m}||^{2}\right\}} \in \mathcal{C}^{N \times 1}$ determines the direction of the precoding vector and $\mathbb{E}\left\{||\mathbf{w}_{k,m}||^{2}\right\}=P_{k, m}^{\rm tx}$.

The data rate perceived by UE $k$ reads as:
\begin{equation}
    R_{k}(\param) = B \log_{2}(1 + \mathrm{SINR}_{k}(\param)).
\end{equation}


\subsection{Network Power Consumption Model}

In the considered system, the power consumption of the entire network depends on the clustering operation matrix $\mathbf{C}$ and includes the power consumption of the CPU and the \acp{AP} as follows:
\begin{equation}\label{eq:net_power_consumption}
    P_{\mathrm{total}}(\param) = \displaystyle\sum_{\substack{m \in \mathcal{M}}}P_{m}^{\rm AP}(\param) + P^{\rm CPU}(\param).
\end{equation}

\noindent
\textbf{AP power consumption.} The power consumption of an AP $m$, denoted by $P_{m}^{\rm AP}(\param)$, consists of the power required to serve connected \acp{UE} and communicate with the CPU via fronthaul links. Following the model proposed in \cite{Machine_Learning_and_Analytical_Power_Consumption_Models_for_5G_Base_Stations}, we mainly focus on the active antenna units (AAUs), which are the most power-consuming components of the AP. In particular, each AAU operates a single carrier, and the power consumption of AP $m$ is given by:

\begin{align}
    {P}_{m}^{\rm AP}(\param) &= P^{\mathrm{AAU,fix}} + P^{\mathrm{FH,fix}} \\\nonumber
    &+ \displaystyle\sum_{\substack{k \in \mathcal{K}}}c_{k,m} \left(\frac{1}{\zeta}P_{k,m}^{\rm tx} + P^{\mathrm{FH, prec}}\right).
\end{align}
Here, $P^{\mathrm{AAU,fix}}$ comprises the power consumed by an AAU to operate including the power consumption of the baseband unit, the transceivers and the power amplifiers (PAs), and $\zeta$ denotes the efficiency of the PAs and antennas. Also, $P^{\mathrm{FH,fix}}$ and $P^{\mathrm{FH, prec}}$ are the fixed and precoding-dependent power consumption of the $m$-th fronthaul link \cite{Sleep_Mode_Strategies_for_Energy_Efficient_Cell_Free_Massive_MIMO_in_5G_Deployments}.

\noindent
\textbf{CPU power consumption.} The power consumption of the CPU, denoted as $P^{\rm CPU}(\param)$, can be expressed as follows, based on the model proposed in \cite{Energy_Efficient_Cell_Free_Massive_MIMO_Through_Sparse_Large_Scale_Fading_Processing}:
\begin{equation}
    P^{\mathrm{CPU}}(\param) = P^{\mathrm{CPU,fix}} + \sum_{\substack{k \in \mathcal{K}}} R_{k}(\param) P^{\mathrm{CPU, enc}},
\end{equation}
where $P_{\mathrm{CPU}}^{\mathrm{fix}}$ and $P_{\mathrm{CPU}}^{\mathrm{enc}}$ are the fixed and encoding-dependent power consumption of the CPU.

\noindent
\textbf{Network energy efficiency.} Thus, given the clustering $\mathbf{C}$ in the downlink scenario, the energy efficiency $\mathrm{EE}_{T}(\rm bits/Joule)$ of the network is computed as:
\begin{equation}
    \label{eq:EE}
    \mathrm{EE}_{T}(\param) = \mathbb{E}\left[ \frac{1}{P_{\rm total}(\param)}\sum_{\substack{k \in \mathcal{K}}} R_{k}(\param) \right],
\end{equation}
where the expectation in Eq. \eqref{eq:EE} is taken \wrt the random traffic requests and channel realizations.


\section{Social user-centric metrics under different sociality regimes}\label{sec:sociality_regimes}
In this work, we are interested in studying how the sociality regimes of the \acp{UE} and \acp{AP} affect the QoS satisfaction for \acp{UE} and energy efficiency per cluster of \acp{AP} serving each UE.

\subsection{Quality of service provisioning}
Let ${R_{k}^{\rm req}}$ be the time-varying traffic request of UE $k$. To assess the UE QoS satisfaction in the association block, we define the following metric as in \cite{sana2021transferable}:
\begin{equation}\label{eq:QoS}
    \rho_{k}(\param) = \min\left(1, \frac{R_{k}(\param)}{R_{k}^{\mathrm{req}}}\right).
\end{equation}

The QoS satisfaction indicator $\rho_{k}(\param)\in[0,1]$ depends on the clustering operation matrix $\mathbf{C}$. In particular, the UE is fully satisfied if $\rho_{k}(\param)=1$. 

The satisfaction of all users depends on the interaction mechanism governing their behavior in the network. Indeed, the full satisfaction of one user may be at the expense of another. Moreover, depending on the criticality of the data rate requirement expressed by a UE, the latter may be willing to accept the degradation of its QoS, thus freeing up radio resources (\eg APs) to improve that of another. To study such behavior, we define $\alpha\in[0,1]$ as the \emph{social factor} of UE and introduce the following \ac{UC} utility function following the approach in \cite{sociality_regime}:
\begin{equation}\label{eq:UE_utility}
    \gamma_{k}^{\alpha}(\param) =  \alpha\rho_{k}(\param) + \frac{1 - \alpha}{K - 1}\sum_{\substack{k' \in \mathcal{K}\backslash k}} \rho_{k'}(\param).
\end{equation}
The social factor $\alpha$ quantifies the degree of cooperation between \acp{UE} in the network. In particular, \acp{UE} are \emph{fully selfish} when $\alpha=1$ and \emph{fully altruistic} when $\alpha=0$. In the \emph{selfish regime}, each UE is only concerned with the level of its own QoS satisfaction, and ignores the QoS satisfaction of other UEs. In \emph{altruistic regime}, the utility of a UE exclusively depends on the QoS satisfaction of other UEs. As a result, a UE may willingly decrease its own QoS to improve that of others. In \emph{egalitarian regime} ($\alpha=\frac{1}{K}$), \acp{UE} share the same utility, here the sum of their levels of QoS satisfaction, which they all contribute to improve.
\begin{proposition}[]
    \label{prop:ue_social_factor}
    The sum of UEs' utilities is independent of the social factor $\alpha$:    
    \begin{equation}\label{eq:total_QoS}
   \Gamma(\param) = \sum_{\substack{k \in \mathcal{K}}} \gamma_{k}^{\alpha}(\param) = \sum_{\substack{k \in \mathcal{K}}} \rho_{k}(\param),
\end{equation}
where $\Gamma(\param)$ denotes the total QoS satisfaction of \acp{UE} in the network.
\end{proposition}

Proposition \ref{prop:ue_social_factor} expresses the fact that the social factor of \acp{UE} $\alpha$ does not affect the total sum of \acp{UE} utilities, enabling a fair comparison between different cooperation regimes among UEs. The proof of Proposition \ref{prop:ue_social_factor} is given in Appendix $\ref{apx:ue_social_factor}$.

\subsection{Energy efficiency per cluster}
Although $\mathrm{EE}_{T}$ in Eq. \eqref{eq:EE} provides an overall measure of the network's energy efficiency, it does not account for the local performance and energy efficiency of individual clusters of \acp{AP} that work together to serve a specific UE. To do so, we propose to define \emph{cluster energy efficiency} $\mathrm{EE}_k$, a \ac{UC} metric capturing how energy-efficient is the cluster of \acp{AP} given a clustering matrix $\mathbf{C}$ to serve a specific UE $k$:
\begin{equation}\label{eq:clusterEE}
    \mathrm{EE}_k=\frac{R_{k}(\param)}{P_{\mathcal{C}_{k}^{\rm UE}}(\param)}, ~\forall  k \in \mathcal{A}^{+}.
\end{equation}
In Eq. \eqref{eq:clusterEE}, $\mathcal{A}^{+}$ denotes the set of associated \acp{UE} and $P_{\mathcal{C}_{k}^{\rm UE}}(\param)$ is the power consumed by the cluster $\mathcal{C}_{k}^{\rm UE}$ to serve UE $k$ such that $P_{\rm total}(\param) = \sum_{k \in \mathcal{K}} P_{\mathcal{C}_{k}^{\rm UE}}(\param) + \left(P^{\mathrm{AAU,fix}} + P^{\mathrm{FH,fix}}\right)(M - |\mathcal{M}^{+}|)$. Here, $\mathcal{M}^{+}$ denotes the set of engaged APs.

A detailed explanation of how we compute $P_{\mathcal{C}_{k}^{\rm UE}}(\param)$ is given in Appendix \ref{apx:clusterPower}.

Similar as for the UEs, we introduce a social factor $\beta$, expressing the willingness of the \acp{AP} to contribute to optimizing the energy efficiency per cluster, and define the following utility:
\begin{equation}\label{eq:cluster_utility}
    \mathcal{E}_k^{\beta}(\param) =  \beta \mathrm{EE}_{k}(\param) + \frac{1 - \beta}{K - 1}\sum_{\substack{k' \in \mathcal{K}\backslash k}} \mathrm{EE}_{k'}(\param)
\end{equation}

For APs, the social factor $\beta$ has the same significance as $\alpha$ in Eq. \eqref{eq:UE_utility} for the UEs. Specifically, when $\beta=1$, the \acp{AP} are \emph{fully selfish}, where each cluster of \acp{AP} only focuses on its energy efficiency. Similarly, when $\beta=0$, they are \emph{fully altruistic}, considering only the benefit of others. An \emph{egalitarian} behavior is adopted when $\beta=\frac{1}{K}$, balancing between individual and collective considerations.

\begin{proposition}[]
\label{prop:ap_social_factor}
    The sum of clusters utilities is independent of the social factor $\beta$:    
    \begin{equation}\label{eq:total_EE}
   \mathrm{EE}(\param) = \sum_{\substack{k \in \mathcal{K}}} \mathcal{E}_k^{\beta}(\param) = \sum_{\substack{k \in \mathcal{K}}} \mathrm{EE}_{k}(\param),
\end{equation}
where $\mathrm{EE}(\param)$ denotes the total energy efficiency of clusters in the network.
\end{proposition}

As in the UE case, Proposition \ref{prop:ap_social_factor} (demonstrated in Appendix \ref{apx:ap_social_factor}) expresses the fact that the social factor of \acp{AP} $\beta$ does not affect the total sum of clusters utilities, allowing a fair comparison between sociality regimes across the clusters of APs.
It is important to note that $\mathrm{EE}_{T}(\param) \neq \mathrm{EE}(\param)$, as the focus is mainly on optimizing the energy efficiency of local clusters rather than the energy efficiency of the network.

\subsection{Network optimization under different sociality regimes}
In this work, we implement a scalable power allocation scheme, where each AP distributes its available transmit power $P_{\max}^{\rm tx}$ to the served \acp{UE} proportionally to their traffic requests:
\begin{equation}\label{eq:power_allocation}
    P_{k,m}^{\rm tx}(\param) = \frac{c_{k,m}R_{k}^{\rm req}}{\displaystyle\sum_{k' \in \mathcal{K}} c_{k',m}R_{k'}^{\rm req}}P_{\max}^{\rm tx}, \forall k \in \mathcal{K}, m \in \mathcal{M}.
\end{equation}
This strategy allows to provide more data rates to \acp{UE} with greater demand and enhance the QoS satisfaction of those experiencing poor channel conditions.
Consequently, the clustering process inherently determines the power allocation strategy\footnote{We leave the joint user association and power allocation to a future work.}. Thus, we only focus on the user clustering problem, to study the impact of the aforementioned sociality regimes. In this context, while \acp{UE} seek to maximize their perceived QoS, \acp{AP} in turn seek to maximize the energy efficiency of the network. These two different interests can often appear contradictory, requiring to identify trade-offs. To do so, let $\mathcal{B}=\{0,1\}^{|\mathcal{K}|\times|\mathcal{M}|}$ denote the set of all possible clustering matrices. We define the following \emph{multi-objective} network utility function:
\begin{align}\label{eq:multi-objective}
    \mathcal{U}^{\alpha,\beta}\colon \mathcal{B} & \longrightarrow(\mathbb{R}^{+})^{2K} \\[0.2ex]\nonumber
  \param & \longmapsto 
    \begin{pmatrix}
        \gamma_{1}^{\alpha}(\param)\\
        \vdots\\
        \gamma_{K}^{\alpha}(\param)\\
        \mathcal{E}_1^{\beta}(\param)\\
        \vdots\\
        \mathcal{E}_K^{\beta}(\param)
    \end{pmatrix}
    =
    \begin{pmatrix}
    \mathcal{U}_1^{\alpha,\beta}(\param)\\
        \vdots\\    
        \mathcal{U}_{K}^{\alpha,\beta}(\param)\\
        \mathcal{U}_{K+1}^{\alpha,\beta}(\param)\\
        \vdots\\
        \mathcal{U}_{2K}^{\alpha,\beta}(\param)
    \end{pmatrix}
\end{align}

Then, we propose the following multi-objective optimization problem:
\begin{align}
\underset{\mathbf{C}\in\mathcal{B}}{\argmax} &  \quad \mathcal{U}_p^{\alpha,\beta}(\param), \quad \forall p = 1,2,\dots, 2K,
 \tag{$\mathcal{P}$} \label{eq:P}\\[0cm]
    	\mathrm{s.t.~}~ & {} c_{k,m} \in \{0,1\},  \qquad \forall k\in \mathcal{K},m\in \mathcal{M}, &\tag{$\mathcal{C}_{1}$} \label{C1}\\
        {}&\sum_{\substack{k \in \mathcal{K}}} c_{k,m} \leq K_{\max}, \qquad   \forall m \in \mathcal{M}, & \tag{$\mathcal{C}_{2}$} \label{C2}\\
        {}&\sum_{\substack{m \in \mathcal{M}}} c_{k,m} \leq M_{\max}, \qquad  \forall k \in \mathcal{K},   \tag{$\mathcal{C}_{3}$} \label{C3}
\end{align}

In problem \eqref{eq:P}, constraint \eqref{C1} indicates that the association variables $c_{k,m}$ are binary. 
Also, \eqref{C2} constrains each AP $m$ to serve at most $K_{\max}$ \acp{UE} due to limited beamforming and processing capability. Finally, constraint \eqref{C3} ensures that each UE $k$ is not associated with more than $M_{\max}$ APs. 

Problem \eqref{eq:P} is a combinatorial, non-convex, and multi-objective optimization problem making it difficult to solve. In particular, there might not exist a feasible solution that maximizes all objective functions simultaneously. Instead, we focus on identifying Pareto optimal solutions: a set of feasible clustering that cannot be improved in any of the objectives functions without degrading at least one of the others.
\begin{definition}[Pareto front]
    A feasible clustering solution $\mathbf{C}\in\mathcal{B}$ is said to dominate another clustering solution $\mathbf{C}'\in\mathcal{B}$ if 
    \begin{enumerate}[leftmargin=*]
        \item $\forall k\in\mathcal{K},\;\gamma_{k}^{\alpha}(\param) \geq \gamma_{k}^{\alpha}(\mathbf{C}')$ and $\mathcal{E}_k^{\beta}(\param) \geq \mathcal{E}_k^{\beta}(\mathbf{C}')$ and
        \item $\exists k\in\mathcal{K}, \;\gamma_{k}^{\alpha}(\param) > \gamma_{k}^{\alpha}(\mathbf{C}')$ or $\mathcal{E}_k^{\beta}(\param) > \mathcal{E}_k^{\beta}(\mathbf{C}')$.
    \end{enumerate}
    A solution $\mathbf{C}^*$ and the corresponding outcome $\mathcal{U}^{\alpha,\beta}(\mathbf{C}^*)$ is \emph{Pareto optimal} if there does not exist another solution that dominates it. The set of Pareto optimal solutions is called the Pareto front.
\end{definition}
Here, the key idea is that social factors implicitly affect the values of $\Gamma(\mathbf{C}^*)$ and $\mathrm{EE}(\mathbf{C}^*)$, guiding the game towards a specific Pareto front. 
However, following Propositions \ref{prop:ue_social_factor} and \ref{prop:ap_social_factor}, $\Gamma(\mathbf{C}^*)$ and $\mathrm{EE}(\mathbf{C}^*)$ are independent of the social factors ensuring a fair comparison between different sociality regimes.

To identify Pareto optimal solutions, we propose two approaches based on a many-to-many matching game \cite{Matching_theory_for_future_wireless_networks_Fundamentals_and_applications} between two sets of players. On the one hand, \acp{UE} seek to improve their perceived QoS, and on the other, clusters of \acp{AP} seek to improve their energy efficiency. 

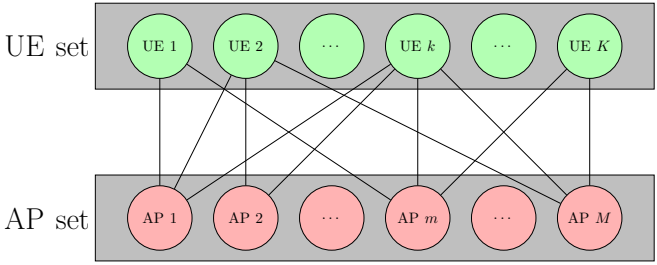
\begin{figure}[!t]
    \centering
    \resizebox{\columnwidth}{!}{
    \begin{tikzpicture}[sbs/.style={draw, circle, fill=red!30, minimum size=1.5cm},
                        ue/.style={draw, circle, fill=green!30, minimum size=1.5cm},
                        set/.style={draw, rectangle, fill=gray!50, minimum width=3cm, minimum height=2cm}]

    \node[set, label=left:\huge{AP set}, minimum width=13cm] (sbsset) at (5,-3) {};
    \node[sbs] (sbs1) at (0,-3) {AP $1$};
    \node[sbs] (sbs2) at (2,-3) {AP $2$};
    \node[sbs] (sbsd1) at (4,-3) {$\dots$};
    \node[sbs] (sbsn) at (6,-3) {AP $m$};
    \node[sbs] (sbsd2) at (8,-3) {$\dots$};
    \node[sbs] (sbsN) at (10,-3) {AP $M$};

    \node[set, label=left:\huge{UE set}, minimum width=13cm] (ueset) at (5,1) {};
    \node[ue] (ue1) at (0,1) {UE $1$};
    \node[ue] (ue2) at (2,1) {UE $2$};
    \node[ue] (ued1) at (4,1) {$\dots$};
    \node[ue] (uek) at (6,1) {UE $k$};
    \node[ue] (ued2) at (8,1) {$\dots$};
    \node[ue] (ueK) at (10,1) {UE $K$};

    \draw (sbs1) -- (ue1);
    \draw (sbs1) -- (ue2);
    \draw (sbs1) -- (uek);
    \draw (sbs2) -- (ue2);
    \draw (sbs2) -- (uek);
    \draw (sbsn) -- (ue1);
    \draw (sbsn) -- (uek);
    \draw (sbsn) -- (ueK);
    \draw (sbsN) -- (ue2);
    \draw (sbsN) -- (uek);
    \draw (sbsN) -- (ueK);
    \end{tikzpicture}
    }
    \vspace*{-0.5cm}
    \caption{Many-to-many matching between \acp{AP} and \acp{UE}.}
    \label{fig:many_to_many_illustration}
\end{figure}
\section{Many-to-many matching game for user association under sociality regimes}\label{sec:M2M}
Matching theory has attracted the interest of wireless network researchers as it enables fully distributed user association with low complexity and fast convergence time \cite{Matching_theory_for_future_wireless_networks_Fundamentals_and_applications}. As shown in Fig. \ref{fig:many_to_many_illustration}, user association can be seen as a many-to-many matching game, involving a dual set of players: \acp{AP} and \acp{UE}.
During this game, each player (AP and UE) starts by constructing its preference list based on local observations. Then, \acp{UE} request connections to \acp{AP} according to their preferences and APs, in their turn, individually decide the \acp{UE} to serve based on their respective preference rankings.

Before formulating the \ac{UCC} as a many-to-many matching game, we first introduce some basic concepts based on two-sided matching theory \cite{Two_sided_matching}.

\subsection{Background on user association matching game concepts}
In the context of a matching game between \acp{UE} and \acp{AP}, preference relations and metrics define how \acp{UE} and \acp{AP} rank each other. \acp{UE} use a preference metric (e.g., perceived SINR) to compare and choose between available \acp{AP}, while \acp{AP} use a preference metric (e.g., channel gain) to rank UEs. These preferences guide the decision-making process in forming clusters, where \acp{UE} aim to connect with their most preferred \acp{AP}, and \acp{AP} select the \acp{UE} they prefer to serve.

Before starting a matching game, the first step is to build preference lists of the players (UEs and APs). Each UE $k$ (resp. AP $m$) constructs its own preference list $\mathcal{P}_{k}^{\rm UE}$ (resp. $\mathcal{P}_{m}^{\rm AP}$) over the set of all \acp{AP} (resp. UEs) in descending order of preferences. For wireless networks, these lists can be constructed through various metrics, such as channel gains, SINRs or data rates of UEs. In our specific scenario, we build the preference lists for both \acp{UE} and \acp{AP} based on channel norms as in \cite{Distributed_user_association_in_B5G_networks_using_early_acceptance_matching_game}:
\begin{equation}\label{eq:channel_fro_norm}
    \vartheta_{k,m}^{\rm UE} = \vartheta_{k,m}^{\rm AP} = ||\mathbf{h}_{k,m}||_{F}, \forall k \in \mathcal{K}, m \in \mathcal{M}, 
\end{equation}
where $\vartheta_{k,m}^{\rm UE}$ (resp. $\vartheta_{k,m}^{\rm AP}$) is the preference value assigned by UE $k$ (resp. AP $m$) to AP $m$ (resp. UE $k$) and the operator $||.||_{F}$ represents the Frobenius norm.

\begin{definition}[Many-to-many matching with externalities]
A many-to-many matching $\mu$ is a mapping from the set $\mathcal{K} \cup \mathcal{M}$ into the set of all subsets of $\mathcal{K} \cup \mathcal{M}$.
The matching process is constrained to:
\begin{enumerate}
    \item $\mu(k) = \mathcal{C}_{k}^{\rm UE}  \subseteq \mathcal{M}$ and $|\mu(k)| \leq M_{\max}$, $\forall k \in \mathcal{K}$,
    \item $\mu(m) = \mathcal{C}_{m}^{\rm AP} \subseteq \mathcal{K}$ and $|\mu(m)| \leq K_{\max}$, $\forall m \in \mathcal{M}$,
    \item $k \in \mu(m) \Leftrightarrow m \in \mu(k)$,
\end{enumerate}
where the notation $|\cdot|$ denotes the cardinality of the corresponding set.
\end{definition}

Conditions (1) and (2) represent the matching process between \acp{AP} and \acp{UE} whereas condition (3) guarantees a mutually accepted match between the set of \acp{UE} and APs. Determining the optimal matching strategy $\mu$, which dictates the cluster of \acp{AP} to serve each UE, is a complex task. A significant difficulty arises because the matching state of each UE $k$ is affected by the matching states of other UEs, due to the mutual interference terms in Eq. \eqref{eq:interference}. This strong interdependence highlights the presence of \textit{externalities} in the clustering process, further complicating this process \cite{Many_to_many_matching_with_externalities_for_device_to_device_communications}.

\begin{algorithm}[!t]
\caption{DA-based Social User Association}
\label{alg:DAG}
\KwData{$\mathcal{P}_{k}^{\rm UE}, \forall k \in \mathcal{K}$}
\KwResult{Matching $\mu$}
\textbf{Initialization}: Initialize empty sets of unassociated \acp{UE} $\mathcal{A}^{-}=\varnothing$ and associated \acp{UE} $\mathcal{A}^{+}=\varnothing$ \;
\textbf{DA game:} \\
\While{$\exists \mathcal{P}_{k}^{\rm UE} \neq \varnothing \hspace*{2mm} {\rm with} \hspace*{2mm} |\mu(k)| < M_{\rm max}$}
{
\For{{\rm UE} $k \in \mathcal{K}$}
{
\If{$|\mu(k)| < M_{\rm max}$}
{
UE $k$ requests association with the first AP in $\mathcal{P}_{k}^{\mathrm{UE}}$ \;
Remove this AP from $\mathcal{P}_{k}^{\mathrm{UE}}$ \;
}
}
\For{{\rm AP} $m \in \mathcal{M}$}
{
Sorts requesting \acp{UE} (including \acp{UE} in its waiting list) in descending order \wrt preference metric \;
Only keeps the first $K_{\max}$ \acp{UE} and rejects the remaining \acp{UE} \;
}
}
Establishing connections based on the final waiting lists of \acp{AP} \;
Add associated \acp{UE} to $\mathcal{A}^{+}$ \;
Add unassociated \acp{UE} to $\mathcal{A}^{-}$ \;
\textbf{Social swap-matching process:}\\
\While{there exists a social swap-blocking pair}
{
    \For{{\rm UE} $k \in \mathcal{A}^{+}$}
    {   \For{{\rm UE} $k' \in \mathcal{A}^{+} \backslash k$}
        {   \If{($k$, $k'$) is a social swap-blocking pair}
        {
                $\mu \leftarrow \mu_{\rm swap}$ \;
                \For{{\rm UE} $i \in \mathcal{A}^{+}$}{
                $\gamma_{i}^{\alpha}(\mu) \leftarrow \gamma_{i}^{\alpha}(\mu_{\rm swap})$\; $\mathcal{E}_i^{\beta}(\mu)\leftarrow \mathcal{E}_i^{\beta}(\mu_{\rm swap})$ \;
                }
        }
            }
    }
}
\end{algorithm}

\subsection{Proposed matching games under sociality regimes}

\subsubsection{Proposed Deferred Acceptance User Association game}

The \ac{DA} algorithm provides a robust framework for creating stable matches between two sets of agents based on their preferences \cite{College_admissions_and_the_stability_of_marriage}. The DA game has demonstrated remarkable versatility, extending its applicability across various domains beyond its initial scope  \cite{Deferred_acceptance_algorithms_History_theory_practice_and_open_questions}. More recently, this framework has been shown to have a promising potential to user association in \ac{CF-MIMO} networks. Therefore, we first propose addressing our user association problem through the DA game.

Alg. \ref{alg:DAG} outlines the procedure of the DA-based solution. Before starting the game, we first build preference lists of UEs. During each iteration, every under-quota UE $k$ ($|\mu(k)| < M_{\max}$) applies to the first AP in its preference list $\mathcal{P}_{k}^{\rm UE}$ and then removes this AP from its $\mathcal{P}_{k}^{\rm UE}$ under constraint (\ref{C3}). Each AP then ranks all new applicants with those already in its waiting list in descending order of preferences according to a preference metric. Then, the AP retains only the top $K_{\rm max}$ \acp{UE} on its waiting list and rejects the remaining UEs. In the DA game, \acp{UE} in the \acp{AP} waiting lists will be associated only after the final iteration of the game, \ie establishing AP-UE connections is deferred until the game concludes. For \acp{UE} that are rejected by all \acp{AP}, they are added to the set of unassociated \acp{UE} $\mathcal{A}^{-}$.

\begin{algorithm}[!t]
\caption{EA-based Social User Association}
\label{alg:EAG}
\KwData{$\mathcal{P}_{m}^{\rm AP}, \mathcal{P}_{k}^{\rm UE}, q_{m}^{\rm AP}, q_{k}^{\rm UE}, \forall k \in \mathcal{K}, m \in \mathcal{M}$}
\KwResult{Matching $\mu$}
\textbf{Initialization}: Set $m_{k}=1$, put all \acp{UE} in a rejection set $\mathcal{R}$, and initialize empty sets of unassociated \acp{UE} $\mathcal{A}^{-}=\varnothing$ and associated \acp{UE} $\mathcal{A}^{+}=\varnothing$ \;

\While{$\exists$ {\rm UE} $k \in \mathcal{R} \cap \mathcal{P}_{m}^{\rm AP}\left(1: q_{m}^{\rm AP}\right)$}{
Try to associate each UE $k \in \mathcal{R}$\;
\eIf{$\mathcal{P}_{k}^{\rm UE} \neq \varnothing$}{
UE $k$ applies to its $m_{k}-th$ preferred AP $m$ (namely AP $m$ with $q_{m}^{\mathrm{AP}} > 0$)\;
\eIf{$k \in \mathcal{P}_{m}^{\rm AP}\left(1: q_{m}^{{\rm AP}}\right)$}{
Associate ${\rm UE}~{k}$ to ${\rm AP}~{m}$ using Alg. \ref{alg:association}\;
Remove UE $k$ from $\mathcal{R}$ and add it to $\mathcal{A}^{+}$\;
}{
$m_{k} \leftarrow m_{k} + 1$ \;
Keep UE $k$ in $\mathcal{R}$ \;
}
}
{
Remove UE $k$ from $\mathcal{R}$ and add it to $\mathcal{A}^{-}$\;
}
}
\If{$\mathcal{R} \neq \varnothing$}{
\For{{\rm UE} $k \in \mathcal{R}$}{
\eIf{$\mathcal{P}_{k}^{\rm UE} \neq \varnothing$}{
Each UE $k \in \mathcal{R}$ will be associated to its first preferred AP $m$ (namely AP $m$ with $q_{m}^{\rm AP} > 0$)\;
Associate ${\rm UE}~{k}$ to ${\rm AP}~{m}$ using Alg. \ref{alg:association}\;
Remove UE $k$ from $\mathcal{R}$ and add it to $\mathcal{A}^{+}$\;
}{
Remove UE $k$ from $\mathcal{R}$ and add it to $\mathcal{A}^{-}$\;
}
}
}
\textit{Cluster evolution process} via Alg. \ref{alg:evol_process} \;
\end{algorithm}

This game not only introduces delays in decision-making process but also does not guarantee an optimal solution to problem \eqref{eq:P}. To address this issue, a \emph{swap-matching process} follows the DA game to improve performance \cite{Many_to_Many_matching_user_association_scheme_in_ultra_dense_millimeter_wave_networks}. This process involves switching AP $m$ associated to UE $k$ with AP $m'$ associated to UE $k'$, if this exchange improves the utility of a player without decreasing the utilities of others.
In this case, the pair of \acp{UE} $(k,k')$ is called a \emph{swap-blocking pair}. To consider the sociality regime of \acp{AP} and UEs, we propose the following definition of a \emph{social} swap-blocking pair.

\begin{definition}[Social swap-blocking pair]\label{def:social-swap}
    The pair of \acp{UE} ($k$,$k'$) is a social swap-blocking pair if $\exists m \in \mu(k)$ and $\exists m' \in \mu(k')$ satisfying
\begin{enumerate}
    \item $\forall i \in \mathcal{A}^{+}$, $\gamma_{i}^{\alpha}(\mu_{\rm swap}) \geq \gamma_{i}^{\alpha}(\mu)$ and $\mathcal{E}_i^{\beta}(\mu_{\rm swap}) \geq \mathcal{E}_i^{\beta}(\mu)$ and \item $\exists i \in \mathcal{A}^{+}$, $\gamma_{i}^{\alpha}(\mu_{\rm swap}) > \gamma_{i}^{\alpha}(\mu)$ or $\mathcal{E}_i^{\beta}(\mu_{\rm swap}) > \mathcal{E}_i^{\beta}(\mu)$,
\end{enumerate}
where $\mu_{\rm swap}$ is the matching strategy obtained after swapping AP $m'$ connected to UE $k'$ with AP $m$ connected to UE $k$.
\end{definition}
Condition (1) enforces that the swap-matching operation should not decrease the utility of each UE and each cluster of APs, while condition (2) guarantees that the swap operation improves the utility of either a UE or cluster of APs.

\subsubsection{Proposed Early Acceptance User Association Game}

The \ac{DA} game, when combined with the swap-matching process, results in prohibitive delays, making the associations between \acp{AP} and \acp{UE} a time-consuming process. This property can be a limiting factor for applications with strict latency requirements which makes DA-based user association impractical for providing low-latency communications. 
To address this issue, we propose a novel matching game called \emph{early acceptance} (EA) \cite{Distributed_user_association_in_B5G_networks_using_early_acceptance_matching_game}. In this game, \acp{AP} immediately decide to accept or reject \acp{UE} at each iteration allowing to accelerate the association process. Furthermore, the EA game not only speeds up the association process but also effectively controls the number of AP-UE associations. Consequently, the EA game reduces the total number of fronthaul connections, unlike the DA game, where the number of associations is usually around the maximum of AP-UE associations, where all \acp{AP} or all \acp{UE} become saturated\footnote{The number of AP-UE connections via DA game is usually around $\min(M K_{\max}, K M_{\max})$.}. With fewer AP-UE connections, fewer messages are exchanged between \acp{UE} and \acp{AP} via fronthaul links, thus, limiting communication overhead and processing delays.
\begin{algorithm}[!t]
    \caption{Association between ${\rm AP}~{m}$ and ${\rm UE}~{k}$}
    \label{alg:association}
    Add AP $m$ to $\mathcal{C}^{\rm UE}_{k}$ and UE $k$ to $\mathcal{C}^{\rm AP}_{m}$\;
    Remove UE $k$ from $\mathcal{P}_{m}^{\rm AP}$\;
    Remove AP $m$ from $\mathcal{P}_{k}^{\rm UE}$\;
    $q_{k}^{\rm UE} \leftarrow q_{k}^{\rm UE}-1$\;
    $q_{m}^{\rm AP} \leftarrow q_{m}^{\rm AP}-1$\;
    \If{$q_{m}^{\rm AP}=0$}{
    Remove AP $m$ from $\mathcal{P}_{k}^{\rm UE}, \forall k \in \mathcal{K}\backslash \mathcal{C}_{m}^{\rm AP}$\;}
    \If{$q_{k}^{\rm UE}=0$}{ 
        Remove UE $k$ from $\mathcal{P}_{m}^{\rm AP},  \forall m \in \mathcal{M}\backslash \mathcal{C}_{k}^{\rm UE}$\;}
    \label{alg:algorithm2}
\end{algorithm}

Alg. \ref{alg:EAG} details our EA-based social user-centric clustering. The proposed algorithm takes as input the preference lists of \acp{AP} ($\mathcal{P}_{m}^{\rm AP}, \forall m \in \mathcal{M}$) and \acp{UE} ($\mathcal{P}_{k}^{\rm UE}, \forall k \in \mathcal{K}$) and the quotas of \acp{AP} ($q_{m}^{\rm AP},  \forall m \in \mathcal{M}$) and \acp{UE} ($q_{k}^{\rm UE}, \forall k \in \mathcal{K}$).
The quota of a UE $k$ (resp. AP $m$) is the number of possible remaining associations it can establish. 

We initialize our algorithm by setting the preference index of all \acp{UE} to one ($m_{k}=1, \forall k \in \mathcal{K}$), forming a rejection set $\mathcal{R}$ containing all \acp{UE}, and creating empty sets of associated \acp{UE} ($\mathcal{A}^{+}=\varnothing$),
and unassociated \acp{UE} ($\mathcal{A}^{-}=\varnothing$). The first step of our algorithm consists in maximizing the number of connected UEs. We start by associating each UE to one AP. At each iteration of the game, each UE $k$ applies to each $m_{k}$-th preferred AP $m$ and it will be immediately accepted if it is among the top-$q_{m}^{\rm AP}$ \acp{UE} in the preference list of the AP $m$. Alg. \ref{alg:association} details the association procedure between UE $k$ and AP $m$ as following: 1) Add AP $m$ to the clustering vector of UE $k$ $\mathcal{C}^{\rm UE}_{k}$ and UE $k$ to the clustering vector of AP $m$ $\mathcal{C}^{\rm AP}_{m}$, 2) UE $k$ is removed from the preference list of AP $m$, the rejection set $\mathcal{R}$ and added to the set of associated \acp{UE} $\mathcal{A}^{+}$, 3) AP $m$ is removed from the preference list of UE $k$, 4) the quota of UE $k$ is updated as $q_{k}^{\rm UE} \leftarrow q_{k}^{\rm UE}-1$ , and 5) the quota of AP $m$ is updated as $q_{m}^{\rm AP} \leftarrow q_{m}^{\rm AP}-1$. When an AP reaches its maximum of quota $K_{\max}$, it sends a broadcast message to the \acp{UE} that it is not serving. Upon receiving this message, the \acp{UE} remove this AP from their preference lists, ensuring that they will no longer apply to this AP.

\begin{algorithm}[!t]
    \caption{Social cluster evolution process}
    \label{alg:evol_process}
    \KwData{ $\mathcal{P}_{m}^{\rm AP}, \mathcal{P}_{k}^{\rm UE}, q_{m}^{\rm AP}, q_{k}^{\rm UE}, \forall k \in \mathcal{K}, m \in \mathcal{M}$, and $\mathcal{A}^{+}$}
    \KwResult{Matching $\mu$}
    \While{there exists a social favorable-association pair}{
    \For{{\rm UE} $k \in \mathcal{A}^{+}$}{
    \If{$\mathcal{P}_{k}^{\rm UE} \neq \varnothing$ and $q_{k}^{\rm UE} > 0$}{
    Set $m_{k}=1$\;
    \While{$m_{k} \leq |\mathcal{P}_{k}^{\rm UE}|$ and there exists no favorable-association pair}{
    UE $k$ applies to its $m_{k}$-th preferred AP (namely AP $m$ with $q_{m}^{\rm AP} > 0$)\;
    \eIf{(${\rm AP}~{m}$, ${\rm UE}~{k}$) is a social favorable-association pair}{
    Associate ${\rm UE}~{k}$ to ${\rm AP}~{m}$ using Alg. \ref{alg:association}\;
    \For{{\rm UE} $i \in \mathcal{A}^{+}$}{
                $\gamma_{i}^{\alpha}(\mu) \leftarrow \gamma_{i}^{\alpha}(\mu_{\rm evolve})$\; $\mathcal{E}_i^{\beta}(\mu)\leftarrow \mathcal{E}_i^{\beta}(\mu_{\rm evolve})$ \;
                }
    }{
    $m_{k} \leftarrow m_{k} + 1$ \;
    }
    }
    }
    }
    }
\end{algorithm}

When the updated preference list of a UE $k$ becomes empty, it is added to the set of unassociated \acp{UE} $\mathcal{A}^{-}$. \acp{UE} that are rejected by all \acp{AP} remain in the set $\mathcal{R}$. For each UE $k \in \mathcal{R}$ (not preferred by any AP), we enforce an association with its first AP $m$ in its updated preference list $\mathcal{P}_{k}^{\rm UE}$, even if UE $k$ is not among the top $q_{m}^{\rm AP}$ \acp{UE} in AP $m$'s preference list $\mathcal{P}_{m}^{\rm AP}$. When all \acp{AP} have exhausted their quotas, the remaining \acp{UE} in $\mathcal{R}$ are added to the set of unassociated \acp{UE} $\mathcal{A}^{-}$. Through these enforced associations, we aim to provide the best possible link quality for the \acp{UE} in $\mathcal{R}$ after prioritizing the \acp{UE} favored by the \acp{AP} in the previous step. This approach ensures that we maximize the number of associated \acp{UE} by fully utilizing the available quotas of APs.

At the end of this \emph{first-matching process}, the matching vector $\mathcal{C}_{k}^{\rm UE}$ for each UE $k\in \mathcal{A}^{+}$ is a singleton. If there are \acp{AP} that do not fully utilize their capacities, \ie there are still opportunities for additional associations between \acp{AP} and UEs, and we propose the \emph{social cluster evolution process} given by Alg. \ref{alg:evol_process}. This process expands the clusters of \acp{AP} serving \acp{UE} by enabling the current matching strategy $\mu$ to evolve into a more beneficial matching, denoted $\mu_{\rm evolve}$. This is achieved through iterative identification of a favorable AP $m$ for each UE $k$, which jointly improves the utilities of all \acp{UE} and their clusters of APs. In this case, the pair $({\rm AP}~m, {\rm UE}~k)$ is called a \emph{social favorable-association pair}. 

\noindent
Similarly to Def. \ref{def:social-swap}, we propose the following definition of a social favorable-association pair, taking into account the sociality regime of both \acp{AP} and UEs.

\begin{figure}[!t]
    \centering
    \resizebox{\columnwidth}{!}{
    \begin{tikzpicture}[font=\footnotesize,thick,>=stealth]

      \node[draw,rectangle] at (0,0) {UE $k$};
      \node[draw,rectangle] at (6,0) {AP $\#$};
      \node[draw,rectangle] at (12,0) {CPU};
      
      \draw[-] (0,-0.5) -- (0,-11);
      \draw[-] (6,-0.5) -- (6,-3.5);
      \draw[dotted] (6,-3.5) -- (6,-4.5);
      \draw[-] (6,-4.5) -- (6,-11);
      \draw[-] (12,-0.5) -- (12,-11);
    
      \draw[->] (0,-1) -- (6,-1) node[midway,above] {Apply to first preferred AP $ m \in \mathcal{P}_{k}^{\rm UE}$};
      \node[draw,rectangle, fill=gray!50] at (6,-2) {Check if UE $k \in \mathcal{P}_{m}^{\rm AP}(1:q_{m}^{\rm AP})$ };
      \draw[->] (6,-3) -- (0,-3) node[midway,above] {if UE $k \notin \mathcal{P}_{m}^{\rm AP}(1:q_{m}^{\rm AP})$, AP $m$ rejects UE $k$};
      \draw[->] (0,-5) -- (6,-5) node[midway,above] {Apply to $m_{k}$-th preferred AP $m'\in \mathcal{P}_{k}^{\rm UE}$};
      \node[draw,rectangle, fill=gray!50] at (6,-6) {Check if UE $k \in \mathcal{P}_{m'}^{\rm AP}(1:q_{m'}^{\rm AP})$};
      \draw[->] (6,-7) -- (12,-7) node[midway,above] {if UE $k \in \mathcal{P}_{m'}^{\rm AP}(1:q_{m'}^{\rm AP})$, };
      \node at (9,-7.25) {request utilities evaluation under $\mu_{\rm evolve}$};
      \node[draw,rectangle, fill=gray!50] at (12,-8) {Evaluate utilities under $\mu_{\rm evolve}$};
      \draw[->] (12,-9) -- (6,-9) node[midway,above] {ACK};
      \draw[->] (6,-10) -- (0,-10) node[midway,above] {if ACK=1, AP $m'$ accepts UE $k$};
      \node at (3,-10.25) {otherwise, it rejects UE $k$};
    \draw[dashed] ($(0,-1)$) -- ++(-1,0);     
    \draw[dashed] ($(0,-3)$) -- ++(-1,0); 
    \draw[dashed] ($(0,-5)$) -- ++(-1,0); 
    \draw[dashed] ($(0,-10)$) -- ++(-1,0); 

    \draw[<->] ($(-0.75,-1)$) -- ($(-0.75,-3)$) node[midway,right] {\footnotesize $t_{\rm rej}$};
     \draw[<->] ($(-0.75,-10)$) -- ($(-0.75,-5)$) node[midway,right] {\footnotesize $t_{\rm eval}$};
    \end{tikzpicture}
}
    \caption{Message sequence chart of the social cluster evolution process.}
    \label{fig:seq_chart}
\end{figure}
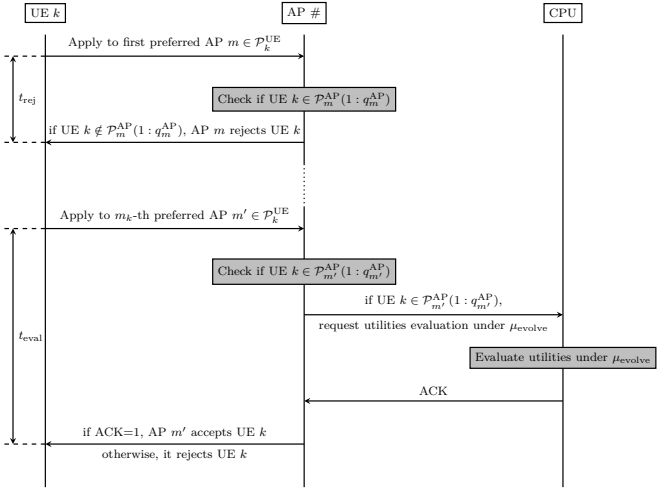

\begin{definition}[Social favorable-association pair]\label{eq:favorable-pair}
     The pair $({\rm AP}~{m}$, ${\rm UE}~{k})$ is a social favorable-association pair if and only if it satisfies
\begin{enumerate}
    \item ${\rm UE}$ $k \in \mathcal{P}_{m}^{\rm AP}\left(1: q_{m}^{\rm AP}\right)$, 
    \item $\forall i \in \mathcal{A}^{+}$, $\gamma_{i}^{\alpha}(\mu_{\rm evolve}) \geq \gamma_{i}^{\alpha}(\mu)$ \\and $\mathcal{E}_i^{\beta}(\mu_{\rm evolve}) \geq \mathcal{E}_i^{\beta}(\mu)$,
     \item $\exists i \in \mathcal{A}^{+}$, $\gamma_{i}^{\alpha}(\mu_{\rm evolve}) > \gamma_{i}^{\alpha}(\mu)$ or $\mathcal{E}_i^{\beta}(\mu_{\rm evolve}) > \mathcal{E}_i^{\beta}(\mu)$.
\end{enumerate} 
Here, $\mu_{\rm evolve}$ is the matching strategy obtained after letting $\mu$ evolve with the new association between AP $m$ and UE $k$.
\end{definition}
Condition (1) indicates that UE $k$ must be among the top-$q_{m}^{\rm AP}$ \acp{UE} in the updated preference list of AP $m$. 
Condition (2) guarantees that the association between AP $m$ and UE $k$ will improve the utility of a UE or a cluster of \acp{AP} without decreasing the other utilities.  Fig. \ref{fig:seq_chart} details the sequence of exchanged messages during the cluster evolution process for a given UE $k$. The UE requests connection to a preferred AP, aiming to form a social favorable association pair. If condition (1) is validated, the AP seeks evaluation of utilities to the CPU to check whether a social favorable association pair can be formed with this UE. The CPU, which has access to all data in the entire network, evaluates in turn the utilities under the new matching $\mu_{\rm evolve}$. It then communicates the evaluation result to the requesting AP. If the new matching $\mu_{\rm evolve}$ demonstrates an improvement without reducing any utility compared to the previous one $\mu$, corresponding to the case when ACK = 1, the AP accepts to serve UE $k$. Conversely, if any utility decreases, the case when ACK = 0, the AP rejects the UE $k$. This results in a message exchange delay, denoted by 
$t_{\rm eval}$, between the UE request and the AP decision when the UE belongs to the AP top-quota preference list.
When an AP becomes saturated and can no longer accept additional UEs, it sends a broadcast message to the \acp{UE} that are not connected to it, and those \acp{UE} in turn remove this AP from their preference lists. Similarly, when a UE is connected to $M_{\max}$ APs, it informs the \acp{AP} that are not associated to it by sending a broadcast message and those \acp{AP} remove this UE from their preference lists. Notably, at each iteration, a UE always reapplies to the first AP in its updated preference list. This reapplication accounts for potential changes in the AP's top-quota set, such as when a previously preferred UE reaches its maximum quota and exits the list, possibly allowing a new UE, previously excluded, to now fall within the top-quota and be associated. The cluster evolution process is repeated until there exists no social favorable-association pair. Hence, we obtain the final matching $\mu$.

\begin{table}[!t]
    \centering
    \small
    \caption{Computational complexity of compared algorithms.}
    \resizebox{\columnwidth}{!}{
\begin{tabular}{|c|c|}
    \hline
    \textbf{User Clustering Alg.}    &  \textbf{Computational Complexity} \\
    \hline
      BC    &  $\mathcal{O}(K  M \log{(M)})$ \\
    \hline
      CS    & $\mathcal{O}(K M)$ \\ 
    \hline
      MD    &  $\mathcal{O}(K  M \log{(M)} + K_{\max} K M)$ \\
    \hline
      DA    &   $\mathcal{O}(K  M \log{(M)} + K M^2 \log(K) + N_{\rm iter}^{\rm SMP}M_{\max}^2 K^3)$ \\
    \hline
      EA  &   $\mathcal{O}(K M \log{(K M)}  +  N_{\rm iter}^{\rm EA} N^{\rm EA} K)$\\
    \hline
    \end{tabular}
    }
    \label{tab:complexity}
\end{table}

\subsection{On the Complexity of the Proposed Matching Games}

The computational complexities of matching games and benchmarks are detailed in Table $\ref{tab:complexity}$.
\subsubsection{Computational complexity of initialization} 
Before initiating the game, it is crucial to construct preference lists. For the DA game setup, each UE sorts \acp{AP} in descending order based on a preference metric, specifically the channel norm in our scenario. This sorting process incurs a computational cost of $\mathcal{O}(K M \log{(M)})$. For the EA game setup, each AP must also build its preference list, ranking \acp{UE} in descending order according to the channel norm. Similarly, the computational cost for constructing the preference lists for \acp{AP} is $\mathcal{O}(M K \log{(K)})$. As a result, the total computational cost for initializing the DA game is $\mathcal{O}(K M \log{(M)})$, while for the EA game, it is $\mathcal{O}(K M \log{(K M)})$.
\subsubsection{DA game versus first-matching process}
In a matching game, a UE requests a connection to an AP, which then decides whether to accept or reject the UE. The execution complexity of this process varies between DA and EA games. In the DA game, each AP, at every iteration, sorts the \acp{UE} requesting association with those already in its waiting list, retaining only the top $K_{\max}$ UEs. This sorting process incurs a computational cost of $\mathcal{O}(M K \log{(K)})$. In contrast, the EA game does not involve a sorting procedure; each AP immediately decides whether to accept or reject the requesting UE, which simplifies the process significantly. Our simulations indicate that the number of iterations required for the DA game approximately equals the number of \acp{AP} ($M$), whereas the EA game typically requires fewer than $M_{\max}$ iterations in most cases. As a result, the overall computational cost for the DA game is $\mathcal{O}(M^{2} K \log{(K)})$, while the EA game incurs a much lower cost of $\mathcal{O}(K)$.

\subsubsection{Swap-matching process versus cluster evolution process}
Here, we assess the computational complexity of swap-blocking process and cluster evolution process under different sociality regimes of \acp{UE} and clusters. In the worse case, where no pair of \acp{UE} shares a common AP, the swap-blocking process involves a computationally expensive operation of $M_{\max}^{2}$ permutations between each pair of UEs, resulting in a total cost of $\mathcal{O}(M_{\max}^2K^{2})$. On the other hand, the cluster evolution process restricts the exploration to $N^{\rm EA}$ iterations, which is upper-bounded by $M_{\max} K$. For each test of a potential social swap-blocking pair or a social favorable association pair, we must ensure that the utilities of the $K$ \acp{UE} and $K$ clusters are not decreased, which costs $\mathcal{O}(K)$. Our numerical results indicate that the number of iterations required by the social swap-blocking process does not exceed $M_{\max}$, while the number of iterations required by the social cluster evolution process, denoted as $N_{\rm iter}^{\rm EA}$, remains below $K_{\max}$. Therefore, in the worst-case scenario, the total computational complexity of the swap-blocking process and cluster evolution process are $\mathcal{O}(N_{\rm iter}^{\rm SMP}M_{\max}^2 K^3)$ and $\mathcal{O}(N_{\rm iter}^{\rm EA} M_{\max} K^2)$, respectively.

\begin{table}[!t]
    \centering
    \caption{Simulation parameters}
            \resizebox{\columnwidth}{!}{
    \begin{tabular}{c c c}
    \hline
    \textbf{Notations}  & \textbf{Parameters}  & \textbf{Values} \\
       \hline
      $f$ &  Carrier frequency  &   3.5 GHz\\
      
      $B$   &  System bandwidth  &  100 MHz \\

      MC & Monte Carlo   &  100 \\    
      
      $N_{\rm MC}$ & Number of simulations per MC  &  100 \\

      $M$ &  Number of \acp{AP}  &   15\\

     $K_{\max}$   &  Maximum number of served \acp{UE} per AP  &   12 \\  
      
     $M_{\max}$      &  Maximum number of associated \acp{AP} per UE  &  6 \\
      
     $N$   &  Number of antennas per AP  &  16 \\
       
     $R_{k}^{\rm req}$  &  UE $k$'s requested throughput  &  $\{100, 300, 500\}$ Mb/s  \\

     $P^{\rm AAU,fix}$    &  Fixed power of each AP  &  40 W \\
     
     $P_{\rm max}^{\rm tx}$    &  Maximum transmission power of each AP  &  250 mW  \\

     $\zeta$  &  Efficiency of the PAs and antennas  &  $40\%$\\

     $P^{\rm FH,fix}$, $P^{\rm FH,prec}$    &  AP fronthaul fixed, precoding-dependent power  &  0.825 W, 0.01 W\\

     $P^{\rm CPU,fix}$, $P^{\rm CPU, enc}$    &  CPU fixed, encoding-dependent power  &  5 W, 0.1 W/Gb/s\\

     $d_{\max}^{\rm LoS}$ &  Maximum distance for LoS component &   30 m  \\

     $N_{p}$  &   Number of multipath components  &  6  \\

     $\Delta$   &   Antenna spacing &   $\lambda / 2$ \\
    $\sigma_{n}^{2}$ &  Noise variance    &   -174 dBm/Hz  \\
       \hline
    \end{tabular}
    }
    \label{tab:simulation_parameters}
\end{table}
\section{Numerical results}\label{sec:results}
In this section, we evaluate the performance in terms of QoS satisfaction and energy efficiency of the proposed algorithms based on DA and EA matching game under different sociality regimes adopted by \acp{UE} and APs. 

\begin{figure*}[!t]
    \centering
    \begin{subfigure}[t]{\columnwidth}
    \centering
        \includegraphics[width=\columnwidth]{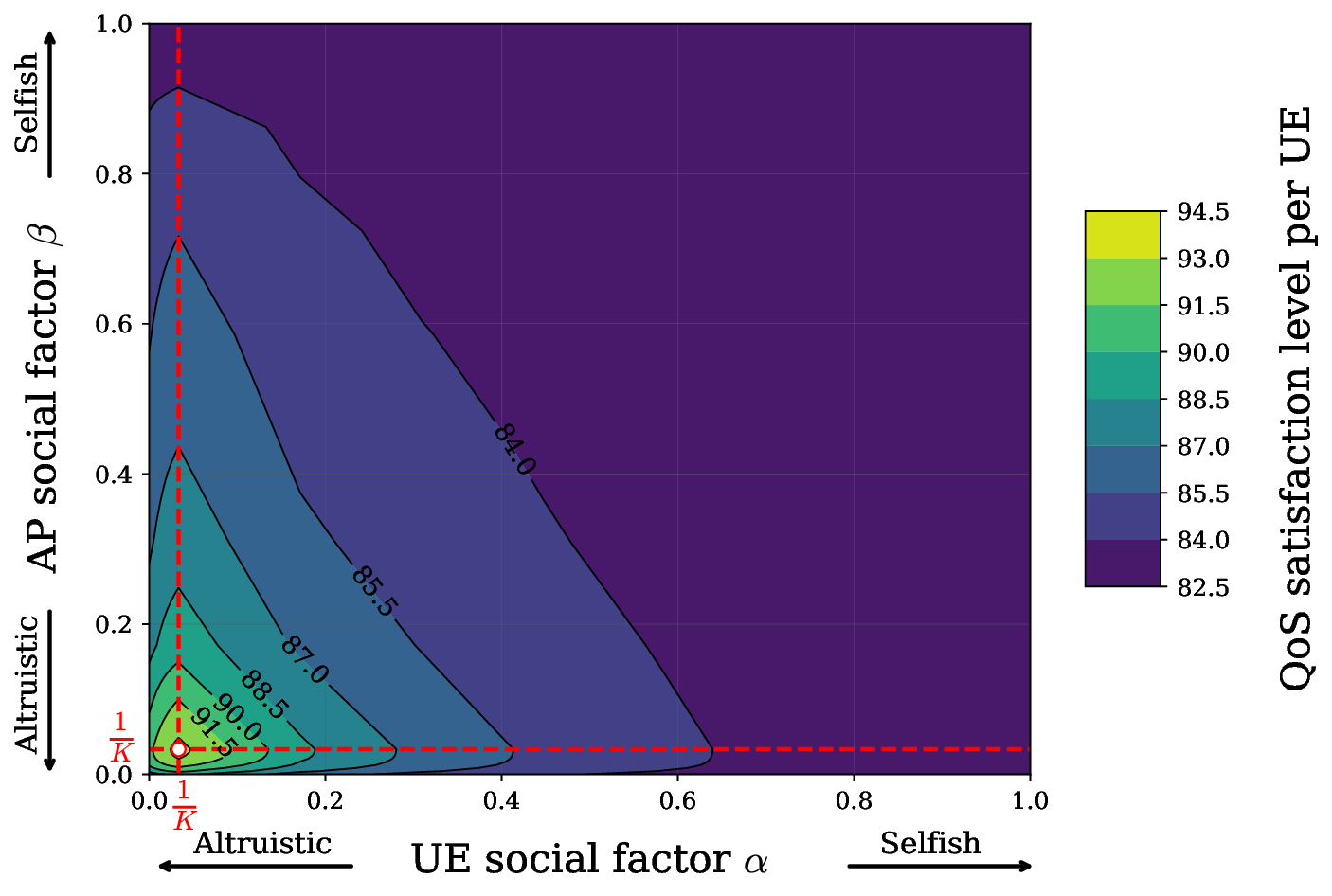}
        \caption{}
        \label{fig:contour_QoS}
    \end{subfigure}
    \hfill
    \begin{subfigure}[t]{\columnwidth}
    \centering
        \includegraphics[width=\columnwidth]{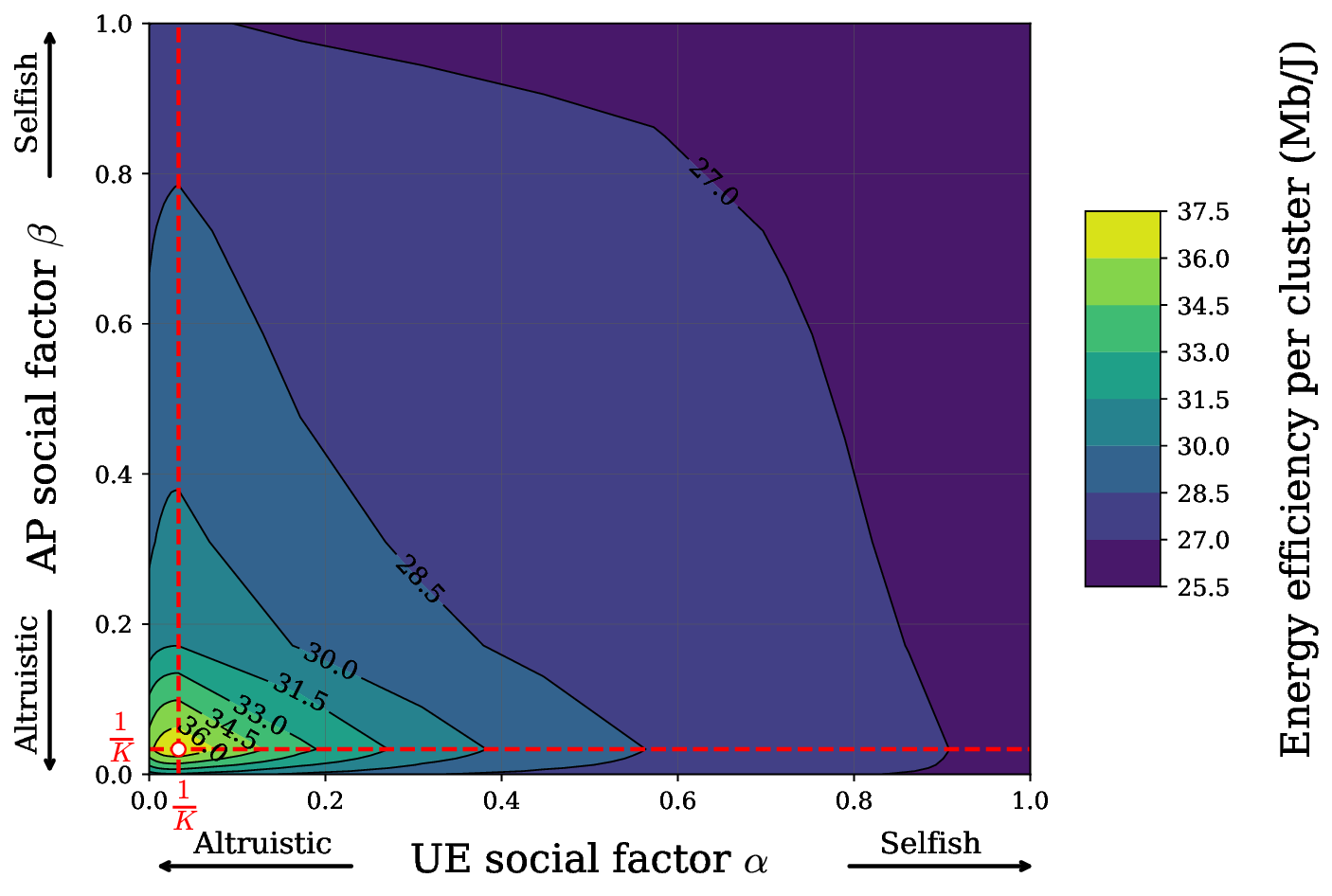}
        \caption{}
        \label{fig:contour_EE_cluster}
    \end{subfigure}
    \hfill
    \begin{subfigure}[t]{\columnwidth}
    \centering
        \includegraphics[width=\columnwidth]{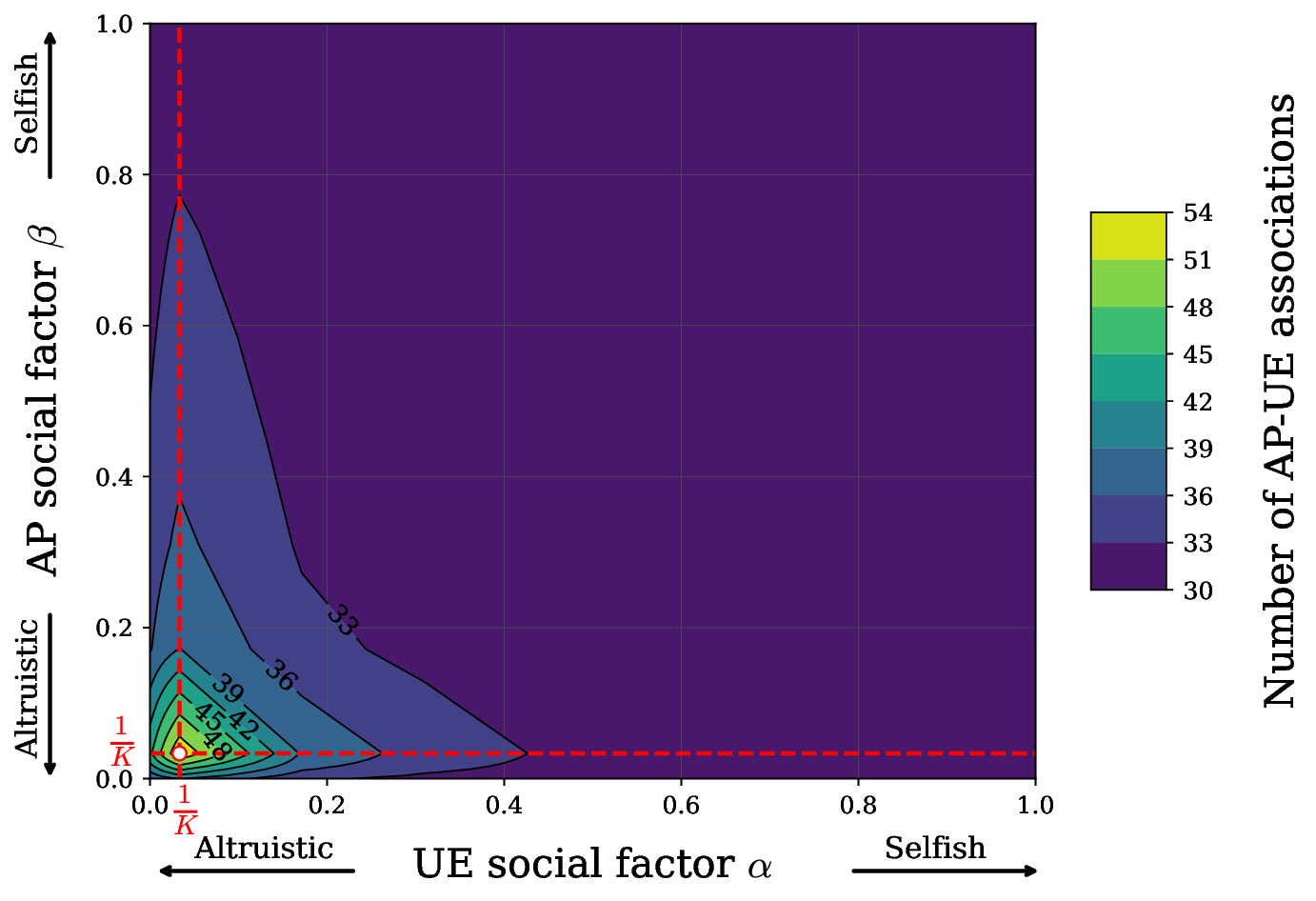}
        \caption{}
        \label{fig:contour_total_associations}
    \end{subfigure}
    \hfill
    \begin{subfigure}[t]{\columnwidth}
    \centering
        \includegraphics[width=\columnwidth]{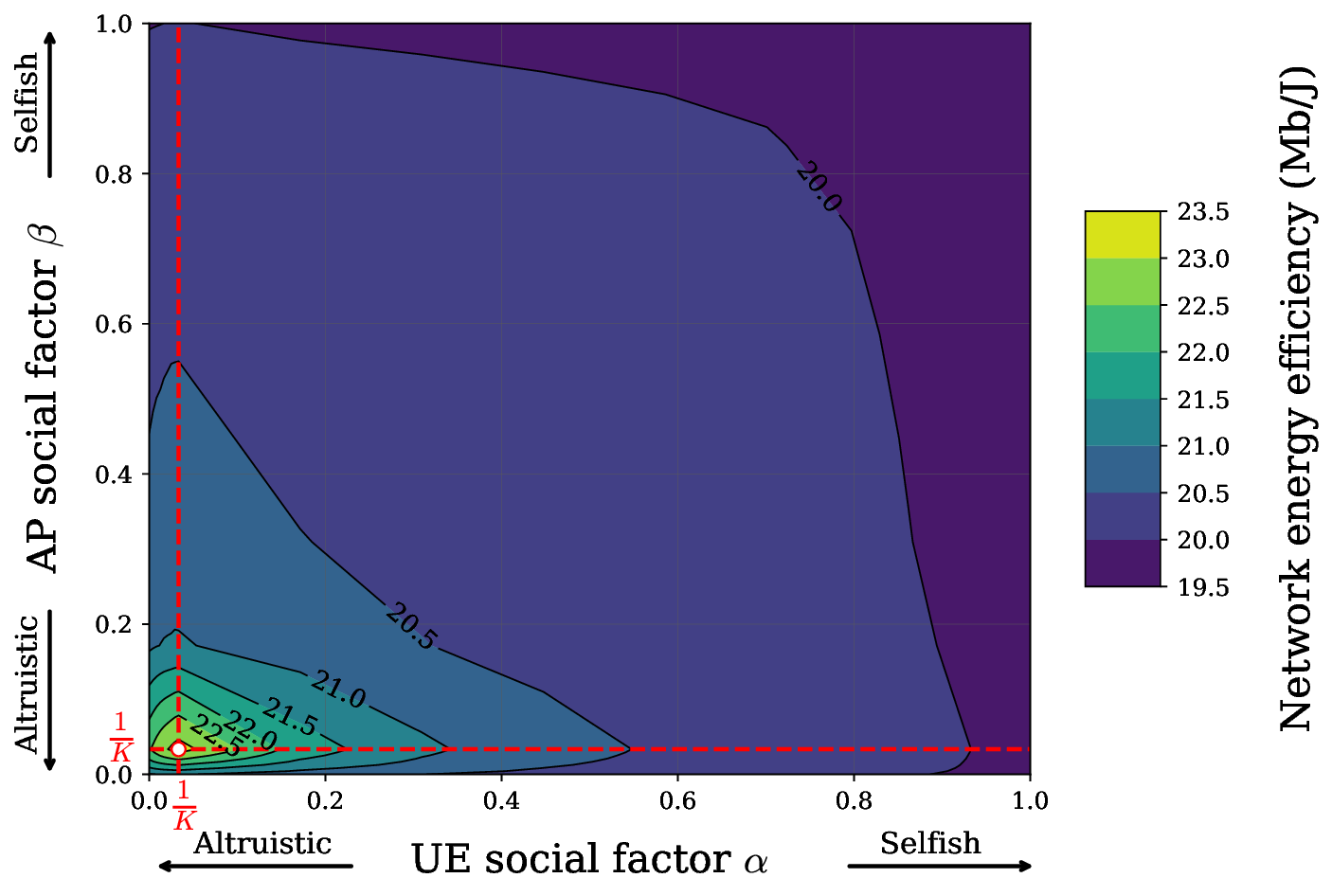}
        \caption{}
        \label{fig:contour_total_EE}
    \end{subfigure}
    \caption{Impact of the social factors $\alpha$ and $\beta$ on the QoS satisfaction per UE and energy efficiency given by EA-based algorithm with $K=30$ and LP-ZF precoding scheme.}
    \label{fig:var_alpha_beta}
\end{figure*}

In our simulations, we consider a \ac{CF-MIMO} network including a CPU, $M = 15$ fixed APs, each equipped  with $N = 16$ antennas that jointly serve $K$ single-antenna \acp{UE} randomly distributed in a region of $97\m \times 36\m$ area\footnote{The same network is simulated using Siradel's simulator, which provides more accurate channel values through \ac{RT} simulations. The results with \ac{RT} channels are evaluated at the end of this section.}. 

The \acp{UE} traffic request follows a Poisson distribution with an intensity randomly sampled from the set $\{100, 300, 500\}\mbs$ during each random deployment. We assume that all \acp{UE} access the full system bandwidth $B$.

Table \ref{tab:simulation_parameters} summarizes the simulation parameters.

We adopt the same channel model as in \cite[Sec. 2.6]{Massive_MIMO_networks_Spectral_energy_and_hardware_efficiency}, where the ($l,i$)-th element of the covariance matrix $\mathbf{R}_{k,m}$ is approximated\footnote{The covariance matrix $\mathbf{R}_{k,m}$ can be approximated when the angular standard deviation (ASD) is small (\eg less than $15^\circ$).} as follows:
\begin{equation}\label{eq:covariance_approximation}
    [R_{k,m}]_{l, i} = \displaystyle\sum_{n=1}^{N_{p}} \frac{e^{j2\pi \Delta(l-i)\sin{(\varphi_{k,m}^{n})}} e^{\frac{-\sigma_{\varphi}^{2}}{2}\left[2\pi \Delta(l-i)\cos{(\varphi_{k,m}^{n})}\right]^{2}}}{N_p}.
\end{equation}
Here, $\varphi_{k,m}^{n} \sim \mathcal{U}\left[\varphi_{k,m} - 40^{\circ}, \varphi_{k,m} + 40^{\circ}\right]$ is the nominal \ac{AoA} for the $n$-th cluster. The multipath components within a cluster have Gaussian-distributed \acp{AoA} around the nominal \ac{AoA}, with an ASD $\sigma_{\varphi}$. 

\begin{figure*}[!t]
    \centering
    \begin{subfigure}[t]{0.66\columnwidth}
    \centering
        \includegraphics[width=\columnwidth]{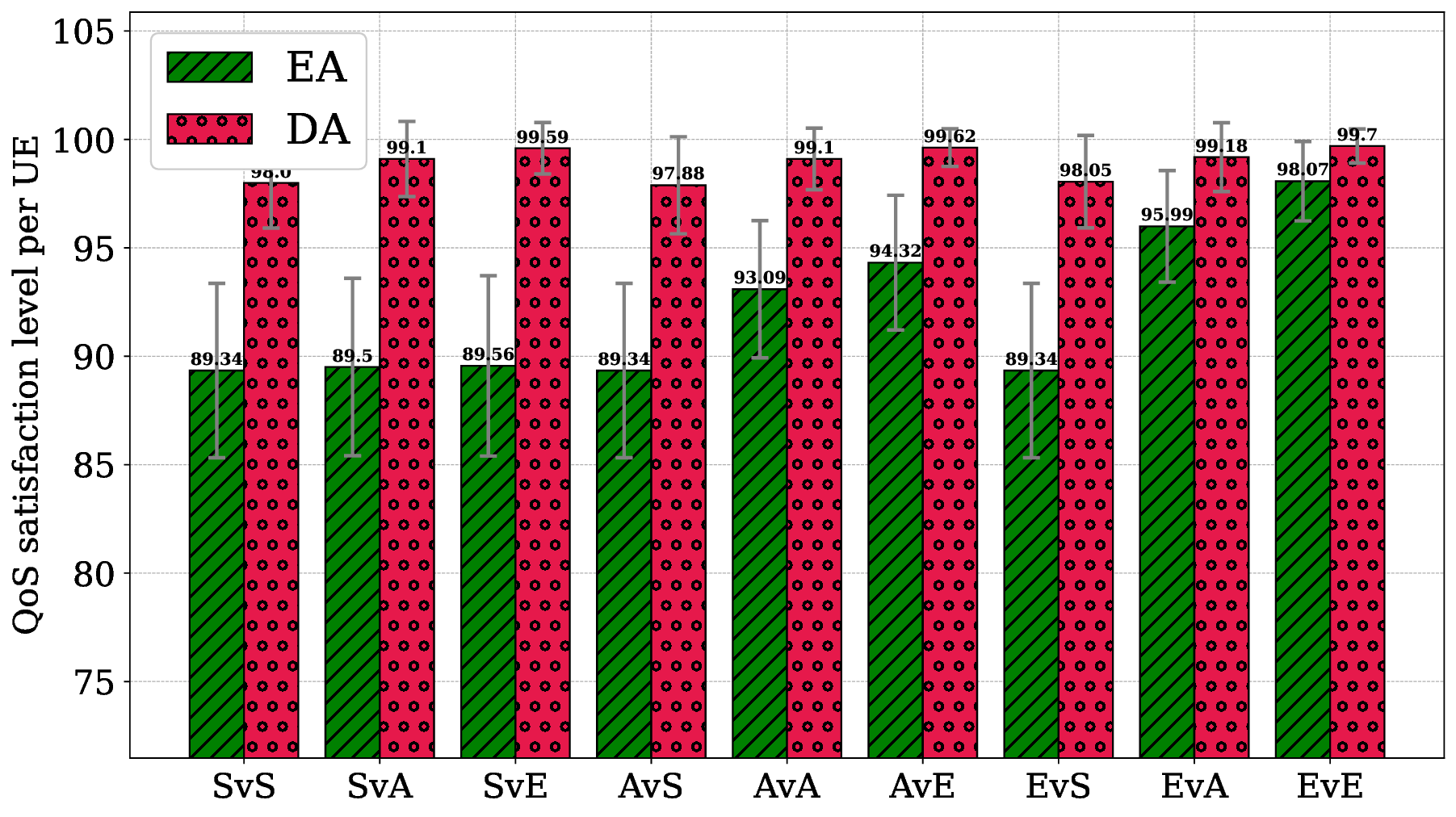}
        \caption{}
        \label{fig:DA_vs_EA_Kappa}
    \end{subfigure}
    \hfill 
    \begin{subfigure}[t]{0.66\columnwidth}
    \centering
        \includegraphics[width=\columnwidth]{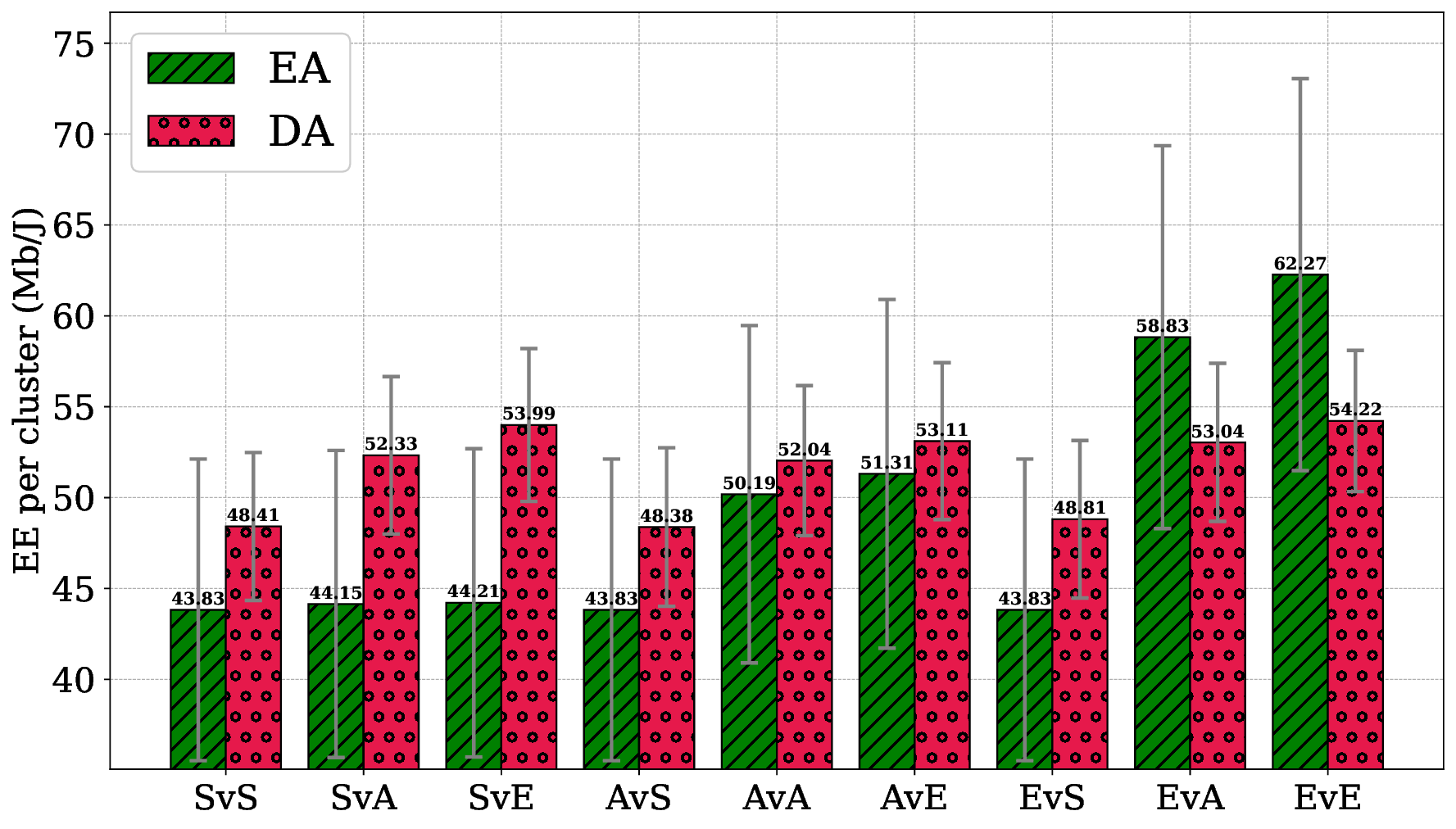}
        \caption{}
        \label{fig:DA_vs_EA_EE_cluster}
    \end{subfigure}
    \hfill
    \begin{subfigure}[t]{0.66\columnwidth}
    \centering
        \includegraphics[width=\columnwidth]{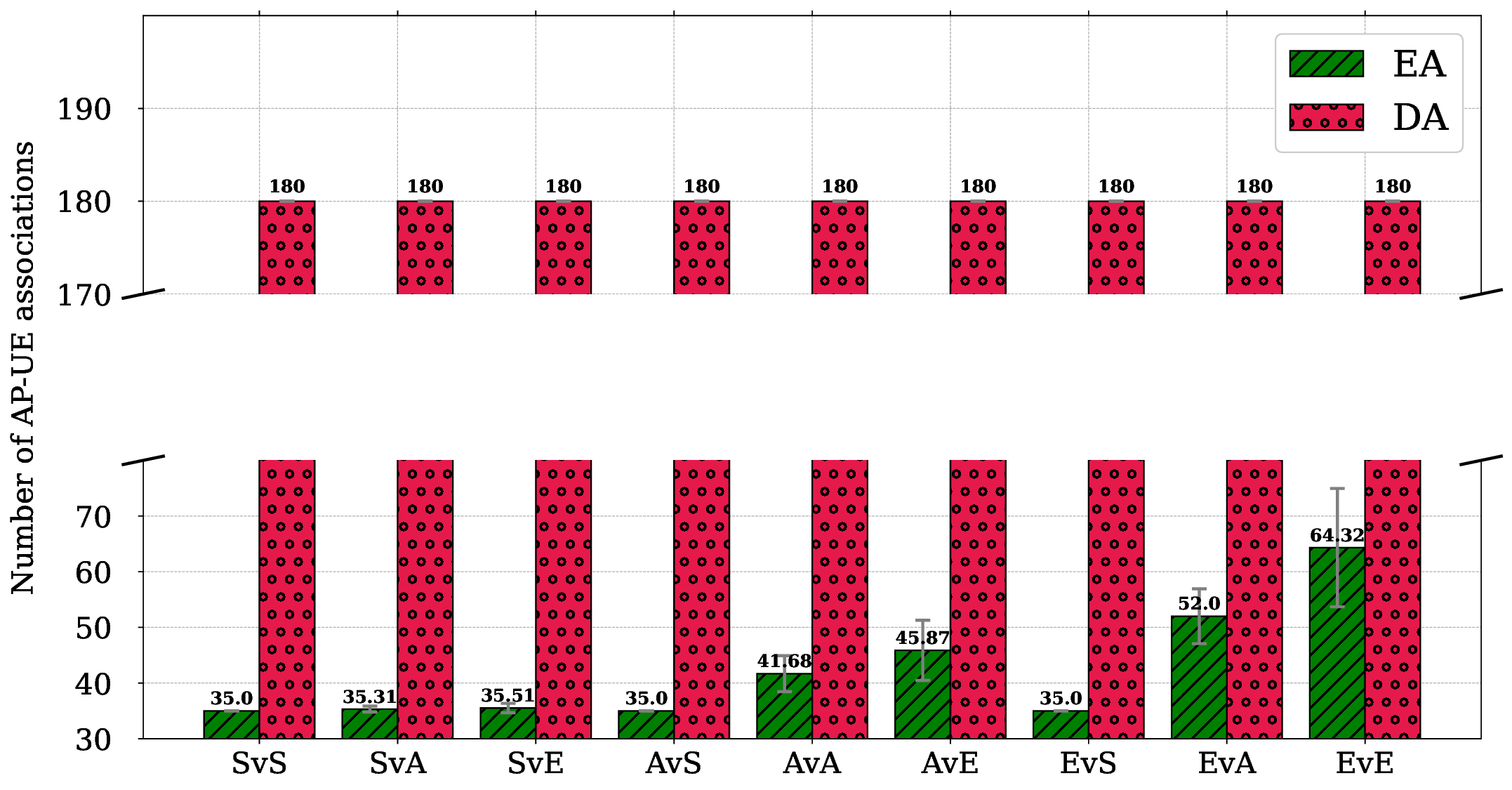}
        \caption{}
        \label{fig:DA_vs_EA_total_associations}
    \end{subfigure}
    \caption{Performance comparison between DA- and EA-based schemes under different combinations of sociality regimes with $K=35$ and LP-ZF precoding scheme.}
    \label{fig:DA_vs_EA_metrics}
\end{figure*}

In our model, the probability of experiencing a LoS component is defined based on the distance between the AP and the UE as:
\begin{equation}\label{eq:prob_LoS}
   {\rm Pr(LoS)} = \left\{
    \begin{array}{ll}
    \frac{d_{\max}^{\rm LoS} - d_{k,m}}{d_{\max}^{\rm LoS}} & \mbox{if} \hspace*{2mm} d_{k,m}\leq d_{\max}^{\rm LoS}, \\
    0 & \mbox{else.}
    \end{array}
    \right.
\end{equation}

When the LoS path exists, the large-scale fading coefficient is modeled (in dB) as:
\begin{equation}\label{eq:path_loss_LoS_dB}
    g_{k,m} = -30.18 - 26\log_{10}(d_{k,m}) + F_{k,m},
\end{equation}
where $F_{k,m} \sim \mathcal{N}(0, \sigma_{s}^{2})$ is the shadowing with $\sigma_{s} = 4$.

If no LoS path exists, the large-scale fading parameter is modeled (in dB) as:
\begin{equation}\label{eq:path_loss_NLoS_dB}
    g_{k,m} = -34.53 - 38\log_{10}(d_{k,m}) + F_{k,m},
\end{equation}
where $F_{k,m} \sim \mathcal{N}(0, \sigma_{s}^{2})$ represents the shadowing with $\sigma_{s} = 10$. The Rician factor is calculated (in dB) as $\kappa_{k,m} = 13 - 0.03d_{k,m}$.

To compute the precoding vectors in distributed way, we adopt the maximum-ratio (MR) and local partial zero-forcing (LP-ZF) as follows:
\begin{itemize}
    \item [$\bullet$] \emph{MR scheme}: A low-complexity precoding that boosts the desired signal but does not mitigate interference among UEs. 
    In this scheme, the precoding vector $\mathbf{v}_{k,m}$ between AP $m$ and UE $k$ that it serves is given by
    \begin{equation}\label{eq:MR_precoding}
        \mathbf{v}_{k,m} = \mathbf{h}_{k,m}.
    \end{equation}
    \item [$\bullet$] \emph{ZF scheme}: A precoding technique designed to suppress interference among \acp{UE} served by the same AP ensuring that signals intended for different \acp{UE} do not interfere with each other. In this scheme, the precoding vector $\mathbf{v}_{k,m}$ is given by
    \begin{equation}\label{eq:ZF_precoding}
        \mathbf{v}_{k,m} = \mathbf{H}_{m}(\mathbf{H}_{m}^H\mathbf{H}_{m})^{-1}e_{i}^{k} = \mathbf{V}_{m} e_{i}^{k},
    \end{equation}
    where $\mathbf{H}_{m} \in \mathcal{C}^{N \times |C_{m}^{\rm AP}|}$ is the channel matrix of AP $m$ with $\mathbf{H}_{m}^H$ denoting the conjugate transpose of $\mathbf{H}_{m}$. The term $e_{i}^{k}$ denotes the $i$-th column corresponding to UE $k$ and $\mathbf{V}_{m}$ represents the precoding vector of AP $m$. 
    The precoder in Eq. \eqref{eq:ZF_precoding} is called ZF because $\mathbf{H}_{m}^H\mathbf{V}_{m} = \mathbf{H}_{m}^H\mathbf{H}_{m} \left(\mathbf{H}_{m}^H\mathbf{H}_{m}\right)^{-1} = \mathbf{I}_{|\mathcal{C}_{m}^{\rm AP}|}$, which implies that there is no interference between \acp{UE} served by the same AP $m$, while the desired signals remain non-zero.
\end{itemize}

\begin{figure*}[!t]
    \centering
    \begin{subfigure}[t]{0.5\columnwidth}
    \centering
        \includegraphics[width=\columnwidth]{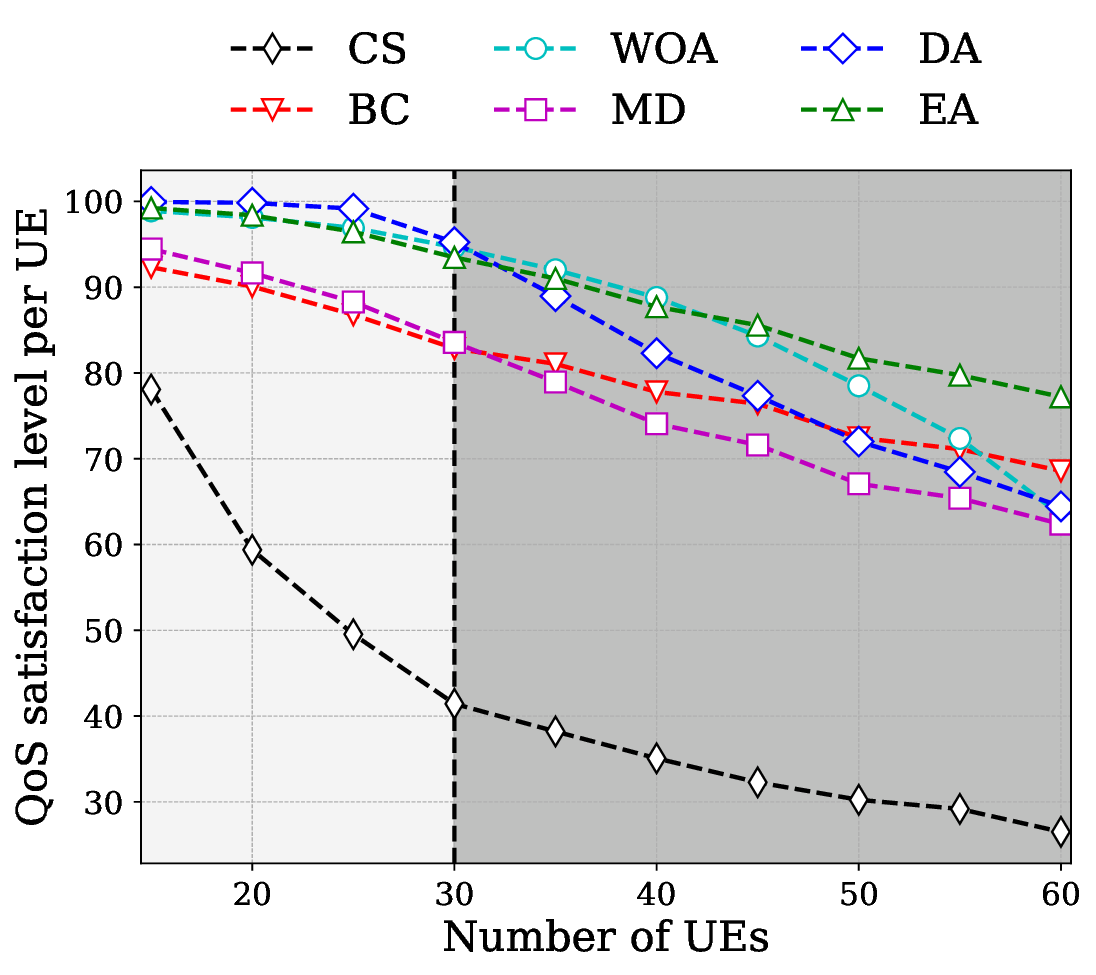}
        \caption{}
        \label{fig:Rice_Kappa}
    \end{subfigure}
    \begin{subfigure}[t]{0.5\columnwidth}
    \centering
        \includegraphics[width=\columnwidth]{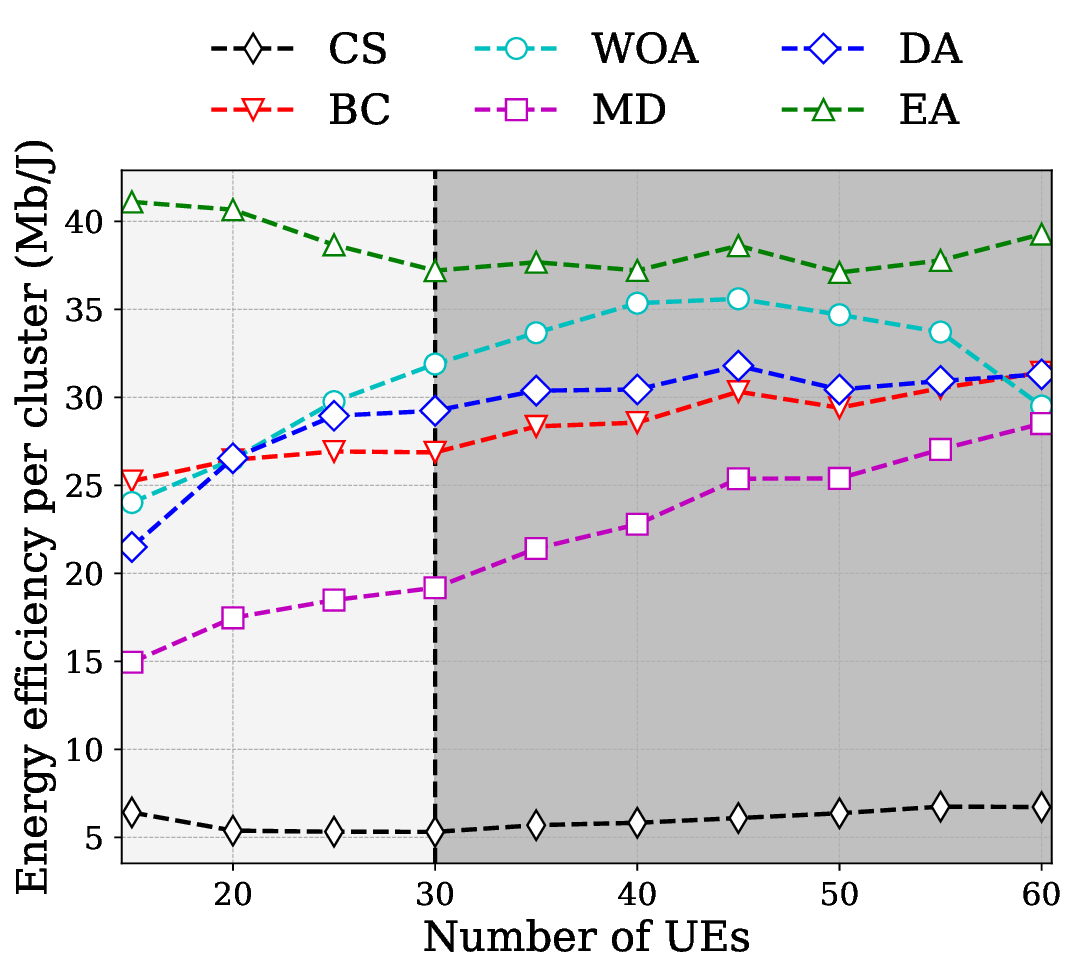}
        \caption{}
        \label{fig:Rice_EE_cluster}
    \end{subfigure}
    \begin{subfigure}[t]{0.5\columnwidth}
    \centering
        \includegraphics[width=\columnwidth]{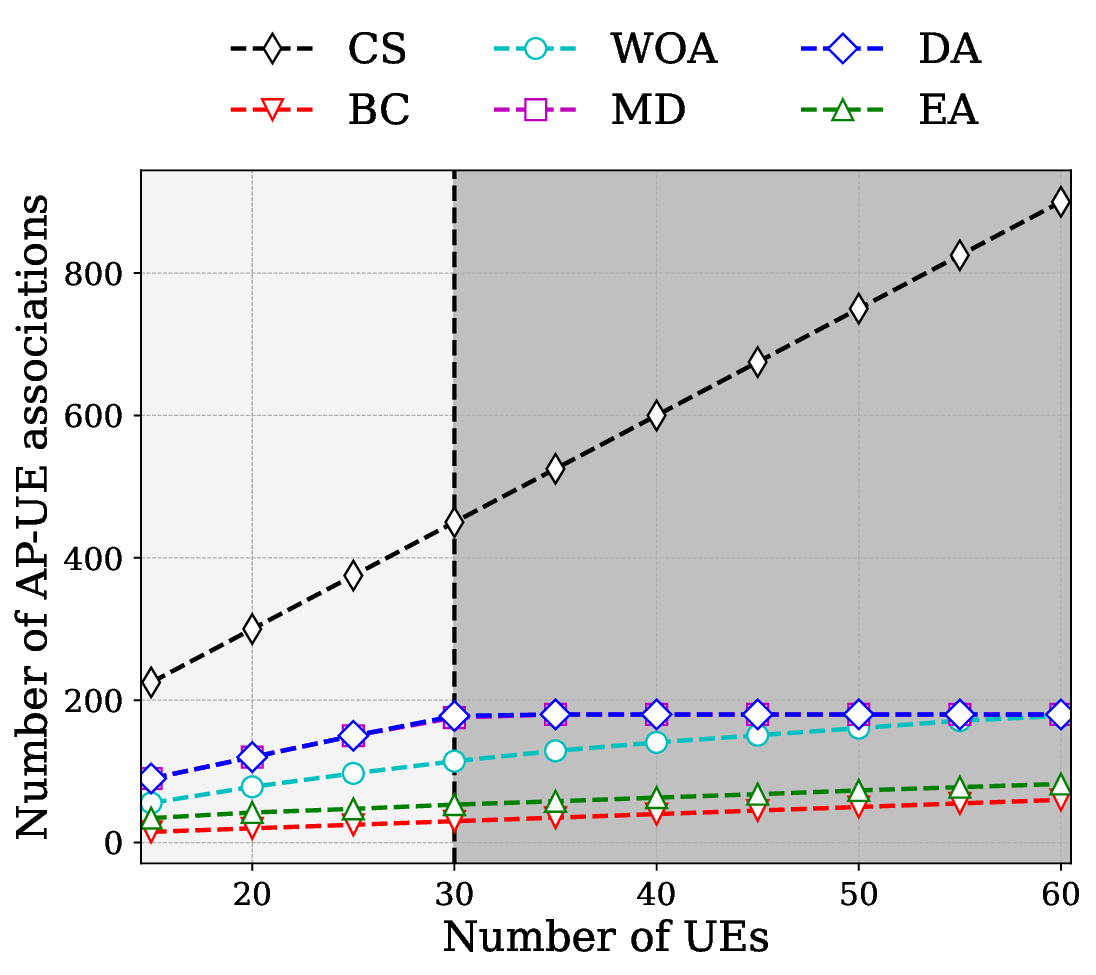}
        \caption{}
        \label{fig:Rice_total_associations}
    \end{subfigure}
    \begin{subfigure}[t]{0.5\columnwidth}
    \centering
        \includegraphics[width=\columnwidth]{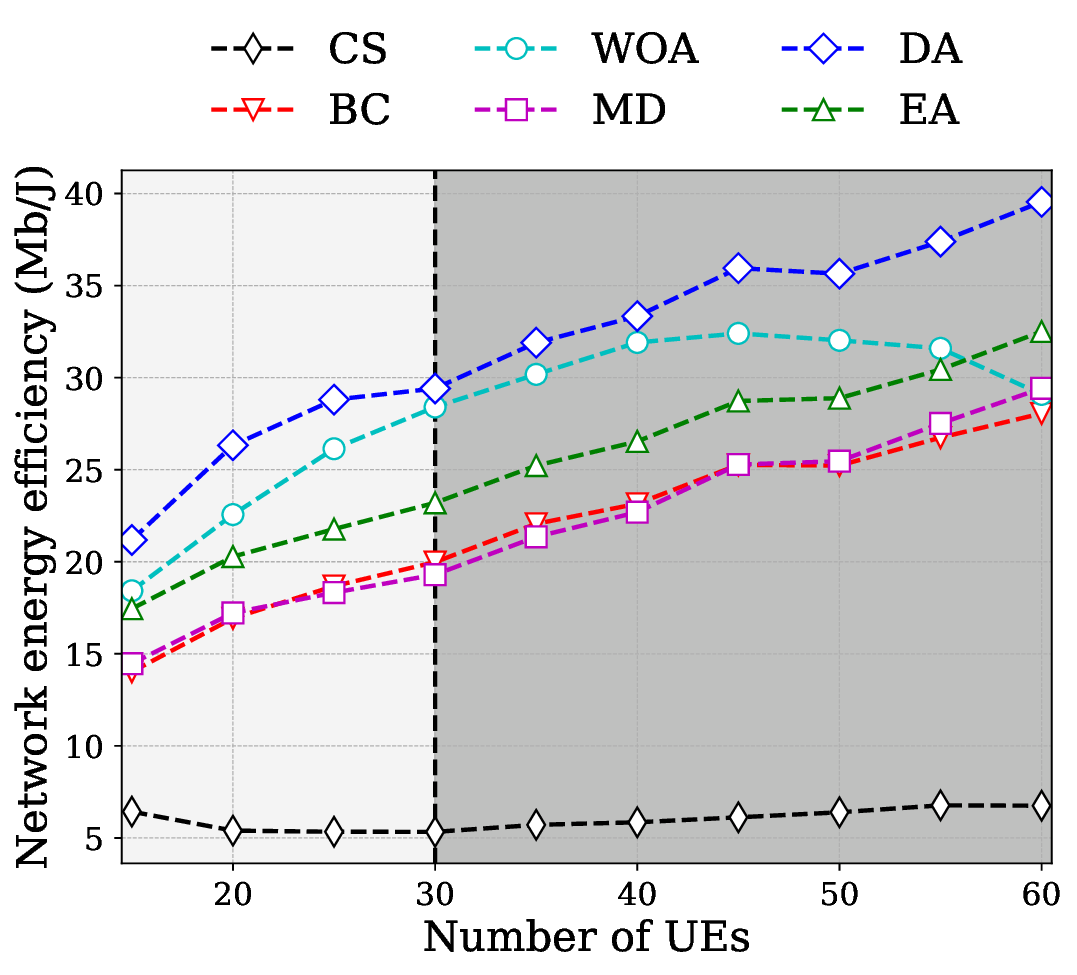}
        \caption{}
        \label{fig:Rice_total_EE}
    \end{subfigure}
    \caption{Effect of different loading scenarios on performance with spatially correlated Rician channels.}
    \label{fig:Rice_diff_K_metrics}
\end{figure*}
\subsection{Evaluation under different degrees of cooperation}
The impact of social factors $\alpha$ and $\beta$ on the perceived QoS and energy efficiency given by the EA-based algorithm is shown in Fig. \ref{fig:var_alpha_beta}. Figs. \ref{fig:contour_QoS} and \ref{fig:contour_EE_cluster} show that the more cooperative the members of a team are, the better the performance in terms of QoS satisfaction per UE and energy efficiency per cluster. We observe that both metrics form equal ranges for the combination $\alpha$ and $\beta$, where typically $\alpha \approx \beta$ for cooperative behavior ($0 \leq \alpha, \beta \leq 0.2$). This shows that both cooperative UEs and cooperative APs equally impact the performance and control the game toward equilibrium. However, when UEs behave selfishly, prioritizing their own QoS satisfaction over the QoS satisfaction of others, the range of performance values influenced by $\alpha$ becomes smaller compared to their depend on $\beta$, where UEs sociality regime dominate that of APs. The difference comes from the way utilities are defined. Specifically, the QoS satisfaction of a UE is calculated as the minimum between 1 and the ratio of its perceived data rate and its requested data rate. Therefore, any UE $k$ that is not fully satisfied ($\rho_{k} < 1$), tends to reject connections between non-serving APs and other UEs, as such connections would generate additional interference and further reduce its utility. 

\fig{fig:contour_total_associations} demonstrates that the cooperative behavior adopted by \acp{UE} and \acp{AP} comes at the cost of more established AP-UE associations in the network, increasing signaling overhead. These additional connections have a significant improvement on QoS satisfaction per UE and both energy efficiency per cluster and energy efficiency of the network (see \fig{fig:contour_total_EE}). 
Thus, these observations highlight the crucial importance of cooperation both among \acp{UE} and within clusters to balance perceived QoS with energy efficiency at local and global scales.
In particular, the highest performance is achieved when \acp{UE} and clusters are egalitarian ($\alpha=\beta=\frac{1}{K}$), outperforming selfish combination up to a $15\%$ improvement in QoS satisfaction per UE, along with enhancements of  $47\%$ in energy efficiency per cluster and $20\%$ in overall network energy efficiency. In fact, adopting an egalitarian behavior ensures equitable treatment when enhancing the QoS satisfaction of each UE and energy efficiency of its serving cluster of \acp{AP}. This approach might decrease the QoS satisfaction or energy efficiency for some \acp{UE} while increasing it for others. This trade-off is crucial for improving collective performance, ensuring fairness, and equitable resource distribution.



\subsection{Performance comparison between DA- and EA-based solutions under different basic sociality regimes combination}
Fig. \ref{fig:DA_vs_EA_metrics} illustrates the comparative performance of DA- and EA-based solutions across varying sociality regimes. The three basic sociality regimes selfish, altruistic and egalitarian gives nine combinations SvS, SvA, SvE, AvS, AvA, AvE and EvS, EvA, EvE. For instance, SvA stands for selfish UEs versus altruistic APs.
Notably, the impact of these regimes on QoS satisfaction and energy efficiency per UE is minor for DA-based algorithm but significantly more pronounced for EA-based approach. In fact, the adoption of selfish behavior blocks the \textit{social cluster evolution process}, limiting the expansion of the APs cluster serving each UE. This occurs because establishing associations in such a scenario induces interference among UEs, thereby reducing their utilities. As a result, most UEs tend to connect to a single preferred AP via the first-matching process, which may not be enough to meet their QoS requirements unlike in the DA-based scheme, where UEs are connected with multiple preferred APs via the DA game. As previously mentioned, Fig. \ref{fig:DA_vs_EA_Kappa} shows that best performance is achieved under egalitarian regime where the two schemes demonstrate comparable levels of QoS satisfaction per UE. However, Fig. \ref{fig:DA_vs_EA_EE_cluster} reveals that the EA-based scheme typically provides superior energy efficiency per cluster under egalitarian regime. This efficiency gain is quantified in Fig. \ref{fig:DA_vs_EA_total_associations}, which shows at least a reduction of approximately $64\%$ in AP-UE connections required by DA-based scheme, significantly reducing the fronthaul signaling and processing delays needed by APs e.g. modulation and beamforming schemes.

The rest of our simulations is done under the egalitarian regime combination adopted by \acp{UE} and their clusters of \acp{AP}, as it provides the optimal balance between QoS satisfaction and energy efficiency per \ac{UE}. For the DA-based approach, we ignore the social swap-matching process due to its prohibitive computational complexity, which arises from the exhaustive search over all potential social swap-blocking pairs, while the performance gains it brings remain marginal.


\subsection{Performance under different loading scenarios} 
\noindent
\textbf{Benchmarks.} We evaluate our EA- and DA-based solutions against benchmarks from the literature under a varying number of UEs in the network:
\begin{itemize}
    \item [$\bullet$] \textbf{Canonical setup (CS)} \cite{Cell_free_massive_MIMO_versus_small_cells}: A non-scalable scheme, where each UE connects to all APs. 
    \item [$\bullet$] \textbf{Best channel (BC)}: Each UE connects to the AP with the highest channel norm, setting a lower bound on the number of possible AP-UE associations in the network.
    \item [$\bullet$] \textbf{Matched decision (MD)} \cite{Matched_decision_AP_selection_for_user_centric_cell-free_massive_MIMO_networks}: Initially, the UE ensures the best connection based on the channel norm. Then, \acp{UE} can expand their clusters by connecting to the unsaturated APs. We modify the proposed algorithm to take into account the limitations on the number of associations $M_{\max}$ that each UE can establish. 
    It sets with the DA algorithm the upper bound on the number of possible AP-UE associations in the network.

    \item [$\bullet$] \textbf{Whale optimization algorithm (WOA)}: Inspired by the bubble-net hunting strategy, WOA has been demonstrated to be an effective heuristic algorithm to solve nonlinear optimization problems \cite{The_whale_optimization_algorithm}. It exploits multiple agents of whales, which collaboratively search the optimal clusters while sharing their searching experiences with each other to improve convergence and solution quality.The action space contains three hunting behaviors, where each whale can encircle prey, bubble-net attack or search randomly. In our simulations, we deploy a population of $50$ whales, where each whale represents a candidate clustering matrix. To initialize each whale,  we firstly sort the \acp{AP} in descending order based on channel norm for each \ac{UE}. Then, each \ac{UE} $k$ is connected to its top $l_k$ preferred \acp{AP}, where $l_k$ is determined as the minimum between a randomly sampled desired cluster size from the set $\{1, \dots, M_{\max}\}$ and the number of available preferred \acp{AP} for UE $k$.
    Finally, we run the WOA algorithm over $N_{ep}$ = 100 epochs to maximize a global utility function $\mathcal{U}_{\rm WOA}$, which combines both the total QoS satisfaction and energy efficiency per clusters as follows
        \begin{equation}\label{eq:WOA_utility}
            \mathcal{U}_{\rm WOA} = \displaystyle\sum_{k\in \mathcal{K}}\rho_{k} + \displaystyle\sum_{k\in \mathcal{K}}\xi_{k}, 
        \end{equation}
        where $\xi_{k} = \min{\left(1, \frac{\rm EE_{k}}{\rm EE_{0}}\right)}$ represents the normalized energy efficiency achieved by cluster $k$ relative to a predefined target ${\rm EE_{0}}$. We choose $\mathrm{EE}_{0} = 35$ Mb/J. Moreover, we adopt an improved version of WOA, where a feedback mechanism is introduced to enhance population diversity and reduce the risk of falling into local optima \cite{A_hybrid_improved_whale_optimization_algorithm}.
\end{itemize}

In particular, an AP $m$ can perform a ZF precoding among the \acp{UE} it serves if $N \geq |\mathcal{C}_{m}^{\rm AP}|$. However, neither CS nor BC ensures that the number of \acp{UE} per AP is less than $N$. To address this, we implement ZF for the first $K_{\max}$ \acp{UE} with the strongest channel norms and MR for the rest \cite{Local_partial_zero_forcing_precoding_for_cell_free_massive_MIMO}. For the remaining algorithms, we usually select $K_{\max} < N$ to perform ZF by each AP to its associated UEs. 
For the DA-based scheme, we simulate only the DA game, excluding the swap-matching process due to its high delay and computational complexity for $K \geq 20$.

\begin{figure*}[!t]
    \centering
    \begin{subfigure}[t]{0.5\columnwidth}
    \centering
        \includegraphics[width=\columnwidth]{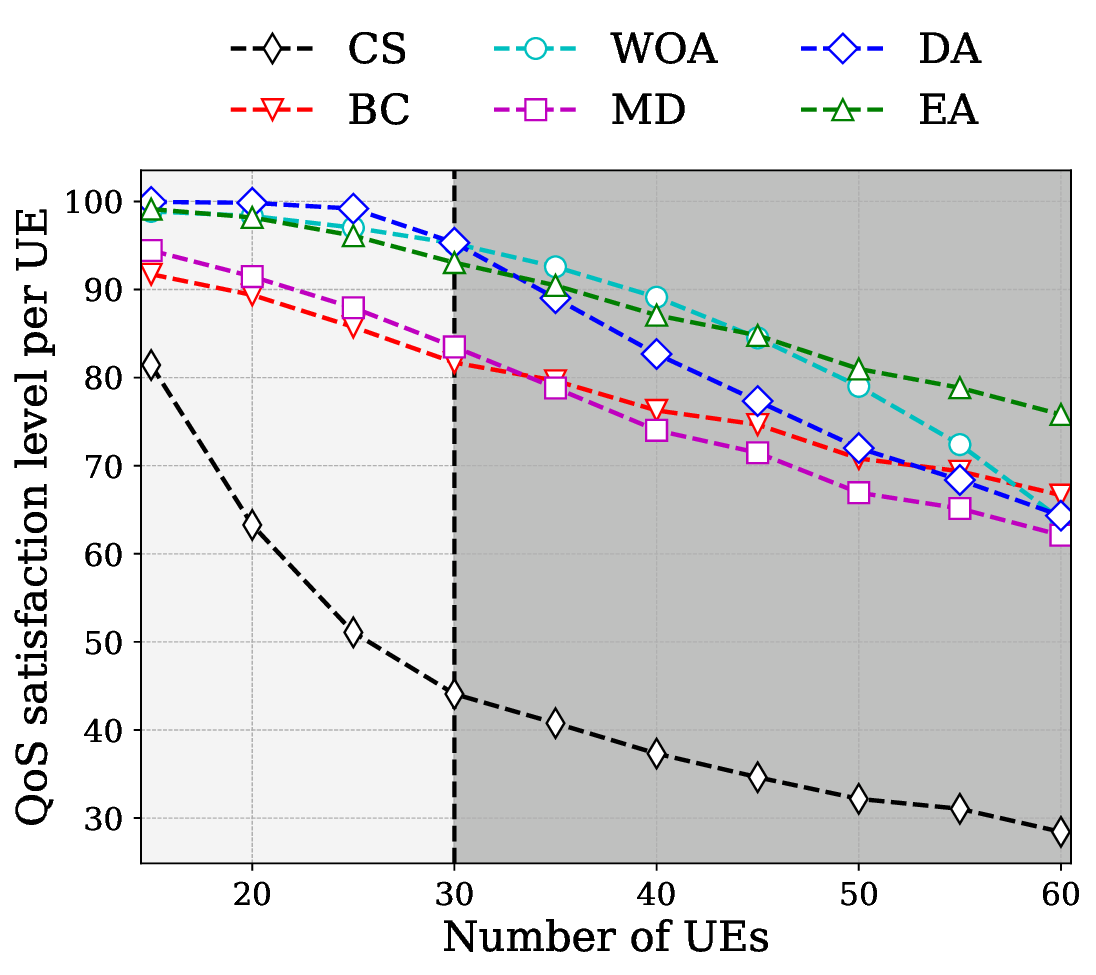}
        \caption{}
        \label{fig:Rayleigh_Kappa}
    \end{subfigure}
    \begin{subfigure}[t]{0.5\columnwidth}
    \centering
        \includegraphics[width=\columnwidth]{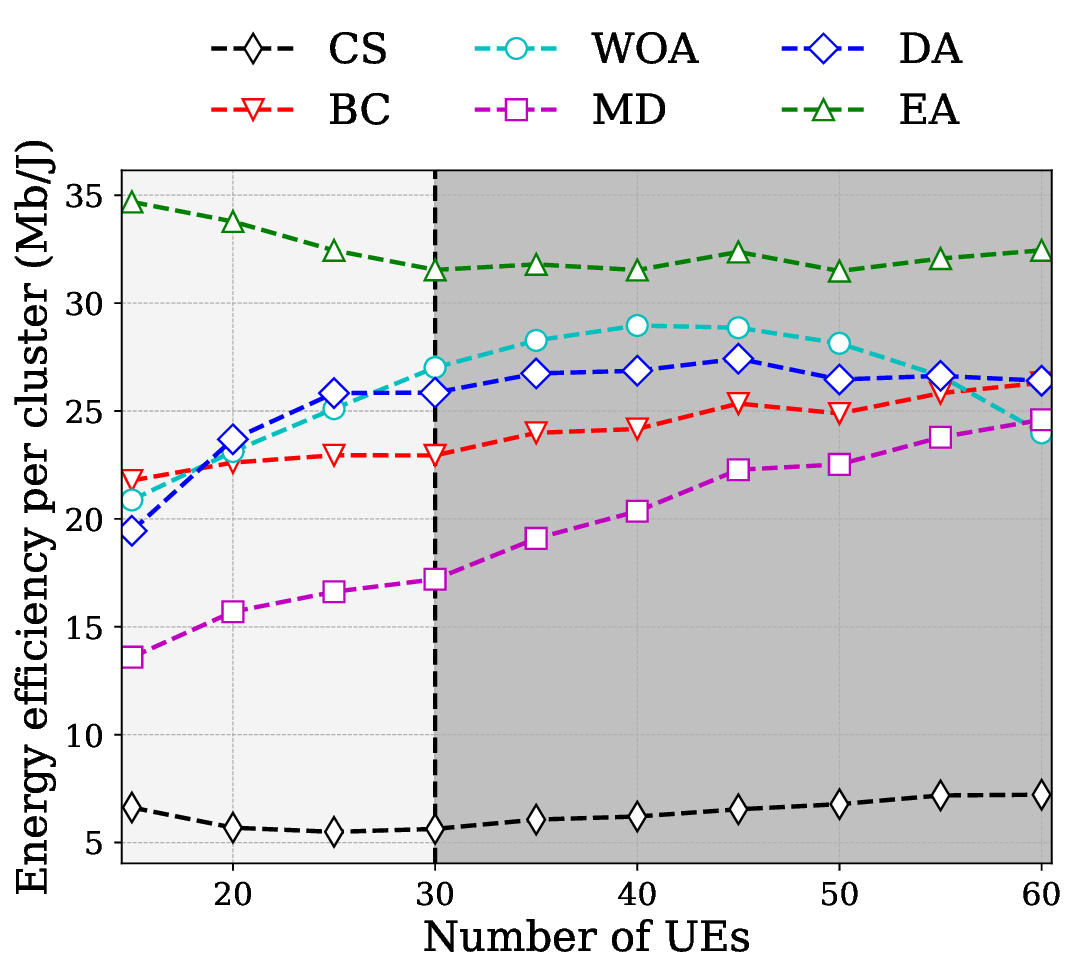}
        \caption{}
        \label{fig:Rayleigh_EE_cluster}
    \end{subfigure}
    \begin{subfigure}[t]{0.5\columnwidth}
    \centering
        \includegraphics[width=\columnwidth]{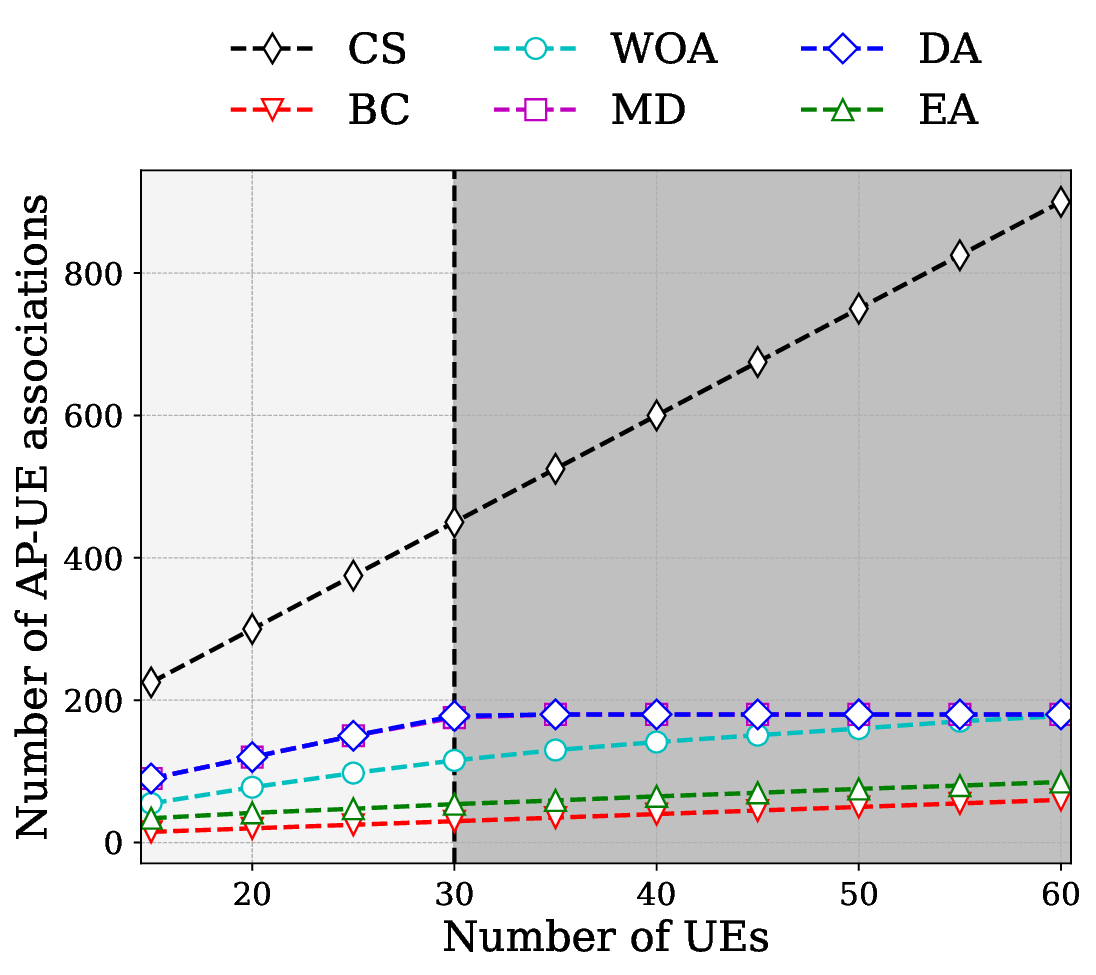}
        \caption{}
        \label{fig:Rayleigh_total_associations}
    \end{subfigure}
    \begin{subfigure}[t]{0.5\columnwidth}
    \centering
        \includegraphics[width=\columnwidth]{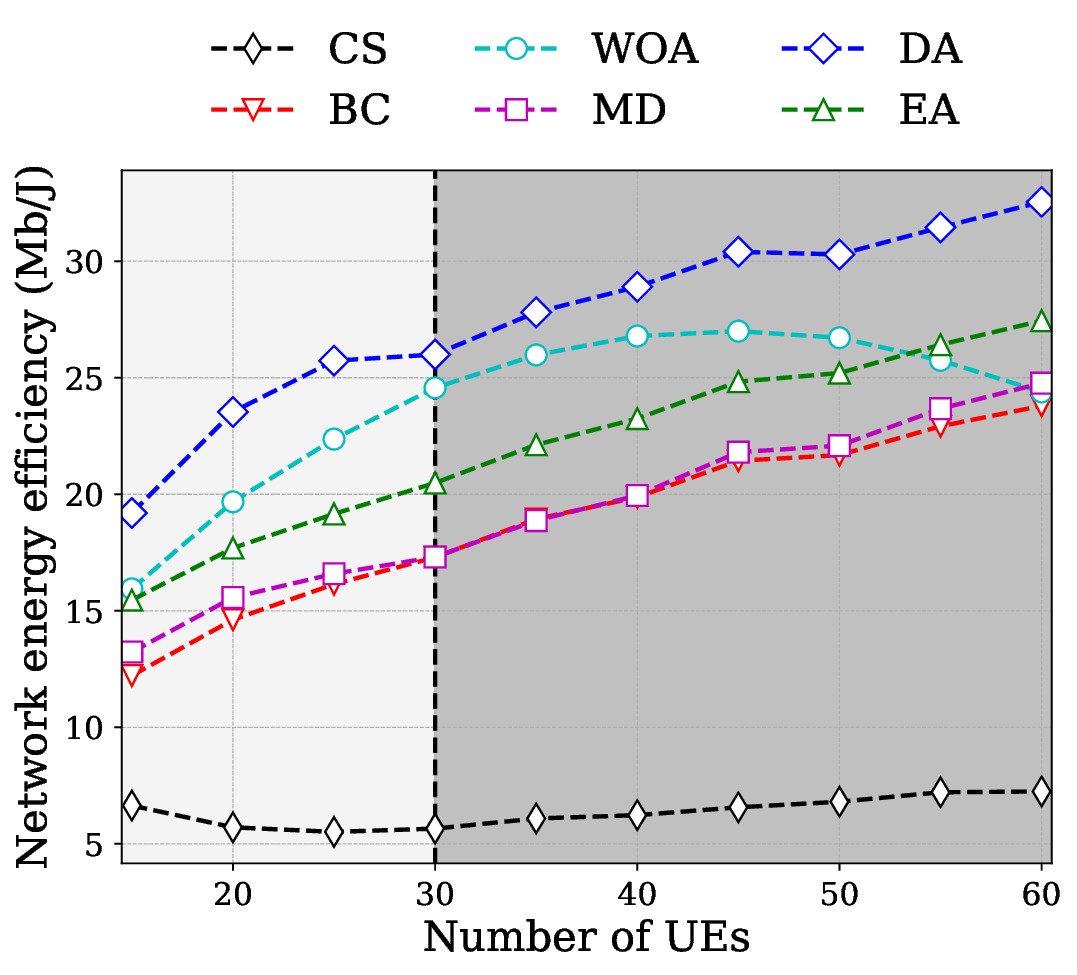}
        \caption{}
        \label{fig:Rayleigh_total_EE}
    \end{subfigure}
    \caption{Effect of different loading scenarios on performance with spatially correlated Rayleigh channels.}
    \label{fig:Rayleigh_diff_K_metrics}
\end{figure*}

In Fig. \ref{fig:Rice_diff_K_metrics}, we evaluate performance of the benchmarks and our proposed approaches under different number of \acp{UE} in the network. The EA-based solution consistently outperforms the benchmark algorithms, achieving the highest energy efficiency per cluster with a satisfactory QoS per UE. Moreover, it substantially reduces the required number of associations between \acp{AP} and UEs, even in overload scenarios ($K \geq 30$), without necessitating AP saturation. In comparison to MD and DA, which adhere to the association constraints of \acp{UE} and \acp{AP}, our algorithm decreases the number of required associations by at least $44\%$ under overload scenarios. Consequently, EA significantly limits fronthaul signaling, as each AP serves a smaller number of UEs. Furthermore, it also limits the computational complexity and processing delay needed for e.g., coding, modulation and channel precoding schemes. 

As shown in Fig. \ref{fig:Rice_total_EE}, the DA algorithm achieves the highest energy efficiency in the network. However, it does not optimally allocate radio resources based on the specific requirements of individual users. This leads to a poor resource allocation strategy, where DA may allocate more and unnecessary resource to a UE that requires less, and vice versa. Consequently, as illustrated in Fig. \ref{fig:Rice_Kappa}, when $K\geq 35$, the EA-based approach outperforms the DA game in terms of perceived QoS. This is due to the ability of the EA-based approach to effectively manage resource distribution among \acp{UE} based on their QoS requirements, unlike the DA game, which becomes less effective in accommodating QoS demands as the number of \acp{UE} in the network increases. Moreover, in particular cases where the number of connections that can be established by APs is smaller than those requested by UEs ($K_{\max}M < M_{\max} K$), the DA algorithm may result in certain UEs being left unassociated, particularly when they are not preferred by any AP, leading to zero data rate for those UEs.

In comparison to DA approach, the MD algorithm demonstrates worse performance, particularly due to its cluster expansion process following the initial intermediate clustering based on channel norms thresholds. Rather than prioritizing the link quality between a given UE and the potential available AP, the algorithm expands clusters based solely on the available quotas of APs. This leads to scenarios where \acp{UE} may be connected to distant APs, thus significantly degrading the overall system performance due to inefficient AP selection.

The WOA algorithm shows good performance across key metrics, outperforming benchmark schemes. This advantage stems from our initialization strategy, which ensures a diverse and high-quality set of initial clusters. During the iterative hunting process over $N_{\rm ep}$ epochs, the whales dynamically improve $\mathcal{U}_{\rm WOA}$, resulting in improved AP-UE connections. Similar to the EA-based approach, Fig. \ref{fig:Rice_total_associations} shows that WOA provides high QoS per user and energy efficiency per cluster without saturating the network, \ie establishing the maximum of AP–UE connections. However, the performance of WOA is notably sensitive to the whale population size and the number of iterations. While increasing these values can yield better solutions, it comes at the cost of greater computational complexity and longer convergence time. 

In comparison to other benchmarks, the CS algorithm demonstrates significant decrease in terms of perceived QoS and local and global energy efficiency as the number of \acp{UE} increases. This is particularly evident in the distributed precoding scheme, which becomes inefficient as the interference between \acp{UE} significantly increases with the growing number of UEs. Consequently, lacking scalability, the CS algorithm results in poorer network performance, making it unsuitable for distributed radio resource allocation in \ac{CF-MIMO} networks.

\begin{figure*}[!t]
    \centering
    \begin{subfigure}[t]{0.5\columnwidth}
    \centering
        \includegraphics[width=\columnwidth]{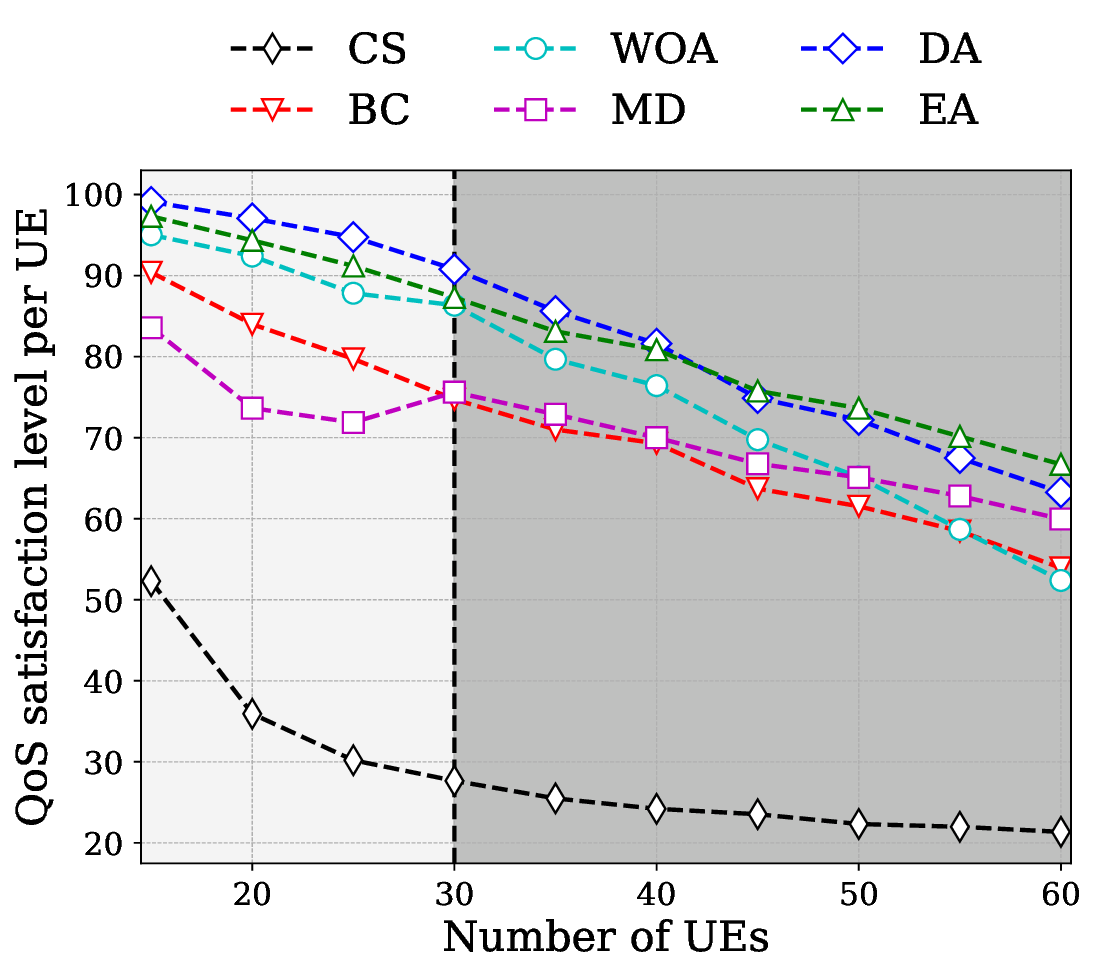}
        \caption{}
        \label{fig:Siradel_Kappa}
    \end{subfigure}
    \begin{subfigure}[t]{0.5\columnwidth}
    \centering
        \includegraphics[width=\columnwidth]{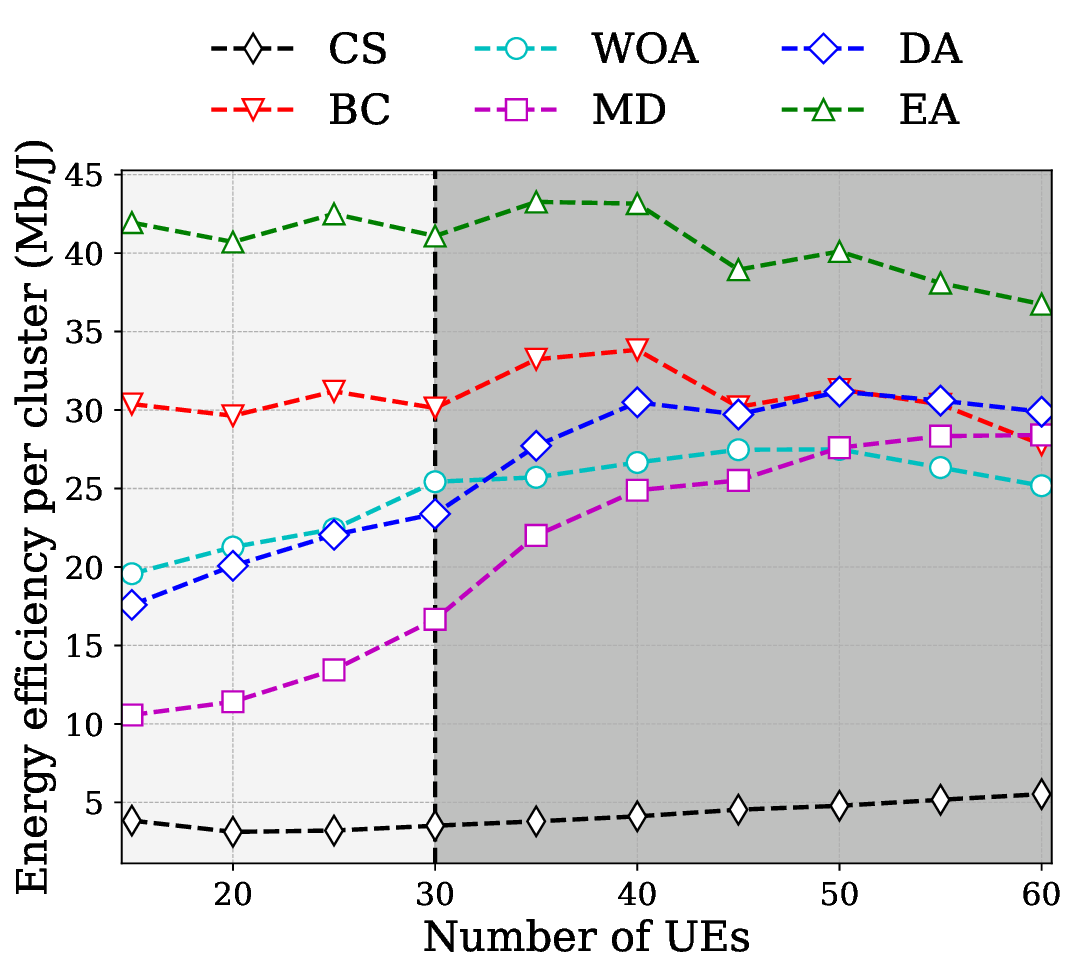}
        \caption{}
        \label{fig:Siradel_EE_cluster}
    \end{subfigure}
    \begin{subfigure}[t]{0.5\columnwidth}
    \centering
        \includegraphics[width=\columnwidth]{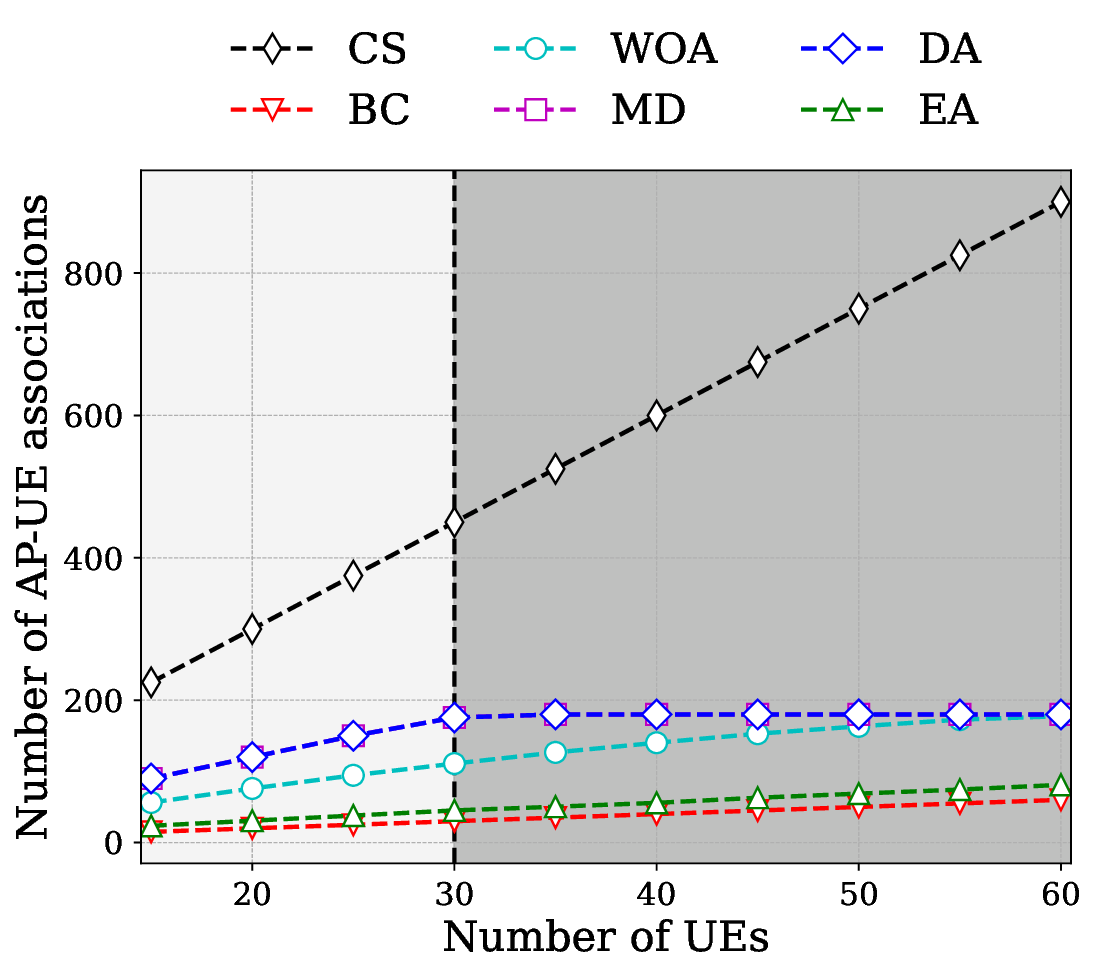}
        \caption{}
        \label{fig:Siradel_total_associations}
    \end{subfigure}
    \begin{subfigure}[t]{0.5\columnwidth}
    \centering
        \includegraphics[width=\columnwidth]{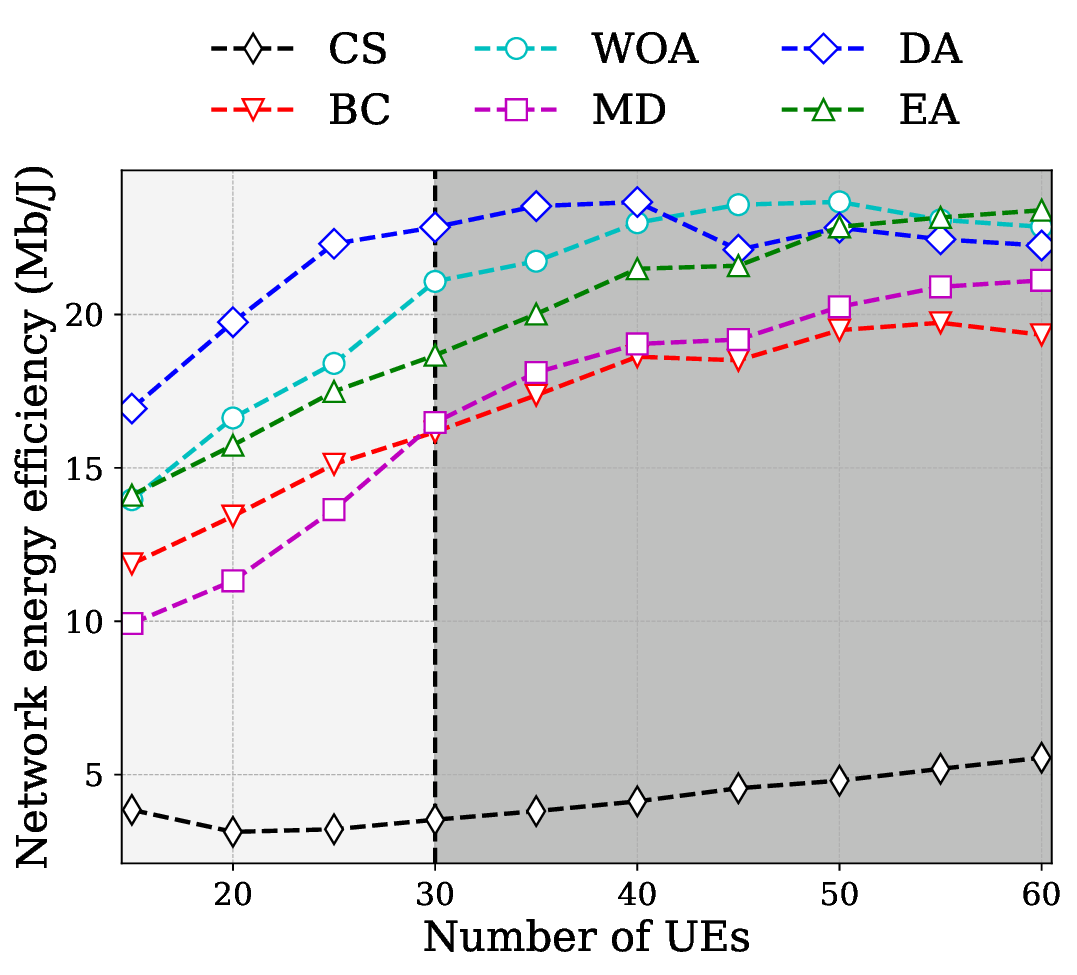}
        \caption{}
        \label{fig:Siradel_total_EE}
    \end{subfigure}
    \caption{Effect of different loading scenarios on performance with \ac{RT} channels.}
    \label{fig:Siradel_diff_K_metrics}
\end{figure*}
\subsection{Performance under different loading scenarios with Spatially correlated Rayleigh channels}
Here, we evaluate the performance of our proposed solutions in a network environment characterized by dense obstacles, which obstruct the direct \ac{LoS} paths between the \acp{AP} and \acp{UE}. This blockage results in spatially correlated Rayleigh channels, where the received signal at each \ac{UE} is dominated by \ac{NLoS} components.
As illustrated in Fig. \ref{fig:Rayleigh_diff_K_metrics}, the QoS satisfaction per \ac{UE} in scenarios with Rayleigh channels is nearly comparable to that observed under Rician channels. However, there is a degradation of approximately $5$ Mb/J in terms of both local and global energy efficiency. This performance gap highlights the critical role played by the \ac{LoS} component in Rician channels. Under a maximum LoS distance $d_{\max}^{\mathrm{LoS}}$ of $30$ meters, although multipath components continue to dominate the received signal, the presence of the direct path significantly improves channel quality due to its lower path loss compared to \ac{NLoS} propagation. Consequently, the absence of a direct \ac{LoS} path in Rayleigh fading channels reduces the perceived data rate per \ac{UE}, causing the observed degradation in energy efficiency performance. In conclusion, the absence of a \ac{LoS} component results in a reduced sum-rate, thus decreasing the energy efficiency. Nonetheless, its impact on QoS satisfaction per \ac{UE}  remains minimal, as the reduction in perceived data rate is negligible relative to the required throughput.

\subsection{Performance under different loading scenarios with ray-tracing channels}
The primary objective of this study is to evaluate the performance of benchmarks and our proposed matching game-based solutions, under more realistic channel realizations. To achieve this, we utilize a channel database from \cite{Siradel24} based on an industrial scenario within a $97 \times 36$ m$^2$ factory. In this setup, $15$ \acp{AP}, each equipped with $16$ antennas, are deployed throughout the space. The database includes $850$ \ac{UE} positions uniformly distributed on a $2$-meter grid across the study area, with each position providing a corresponding  deterministic channel vector generated through \ac{RT} simulations.

A key observation, as illustrated in Fig. \ref{fig:Siradel_diff_K_metrics}, is the degradation in performance when using RT channels. In particular, the RT channels in the proposed scenario are dominated by the LoS component, which limits channel diversity. This results in increased spatial correlation between channels of UEs, thereby reducing the effectiveness of local zero-forcing (ZF), which better separates user channels in multipath propagation with greater spatial diversity. In terms of performance, both RT and stochastic models (Rice and Rayleigh) provide similar qualitative behavior. However, the stochastic model tends to be optimistic\footnote{The same observations were made in \cite{Siradel24}.}, which can lead to an overestimation of network performance especially in terms of interference mitigation compared to more realistic RT-based evaluations, where RT model provides accurate predictions by considering the network's geometry, material properties and polarization diversity. Under RT channels, simulations show that our EA-based solution provides a QoS for each UE close to the QoS provided by DA, while also achieving higher energy efficiency per cluster. These results highlight the robustness of our approach, even under more challenging channel conditions.

\begin{figure}[!t]
    \centering
       \begin{subfigure}[t]{0.49\columnwidth}
        \centering
             \includegraphics[width=\columnwidth]{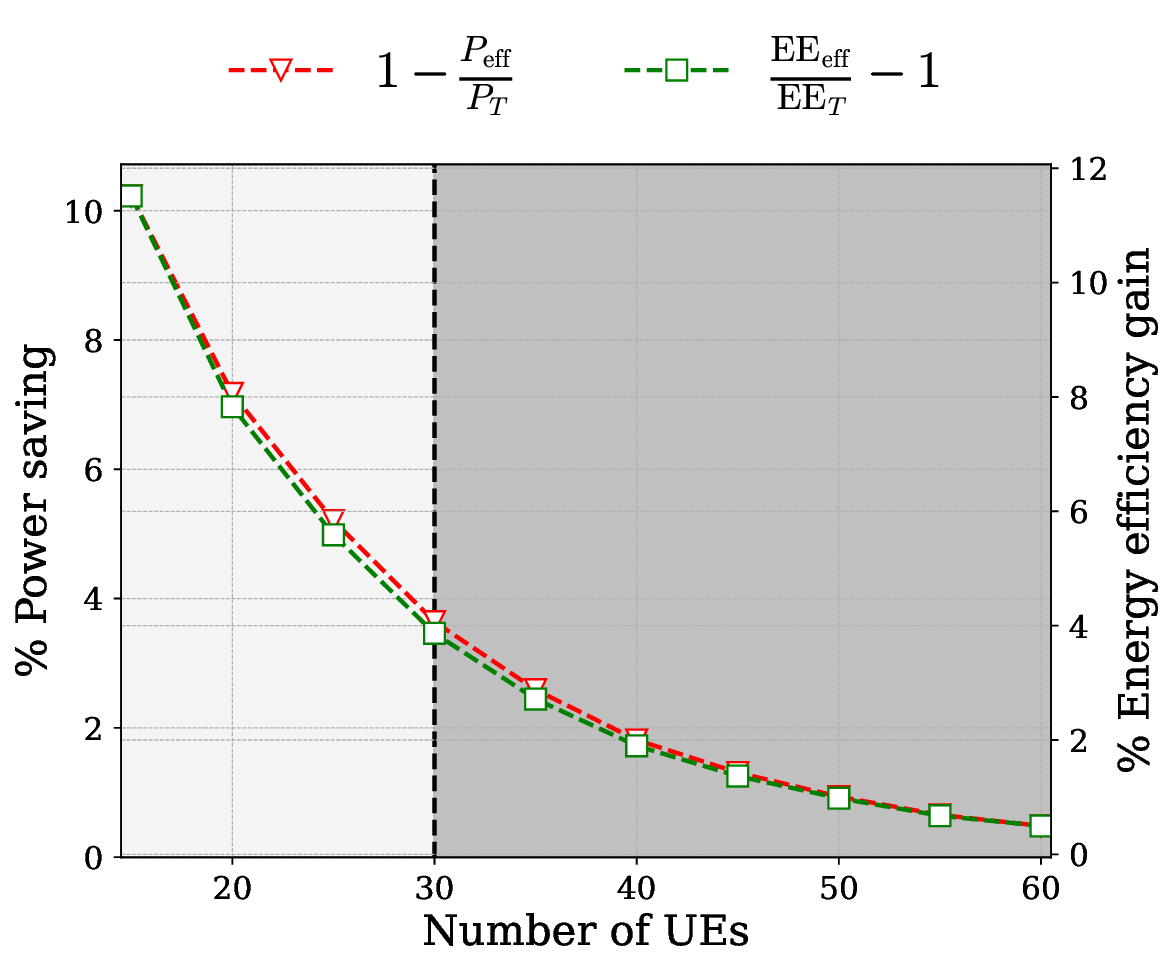}
             \caption{}
            \label{fig:Rice_P_EE_vs_effective_P_EE_EA}
    \end{subfigure}
    \hfill
    \begin{subfigure}[t]{0.49\columnwidth}
        \centering
             \includegraphics[width=\columnwidth]{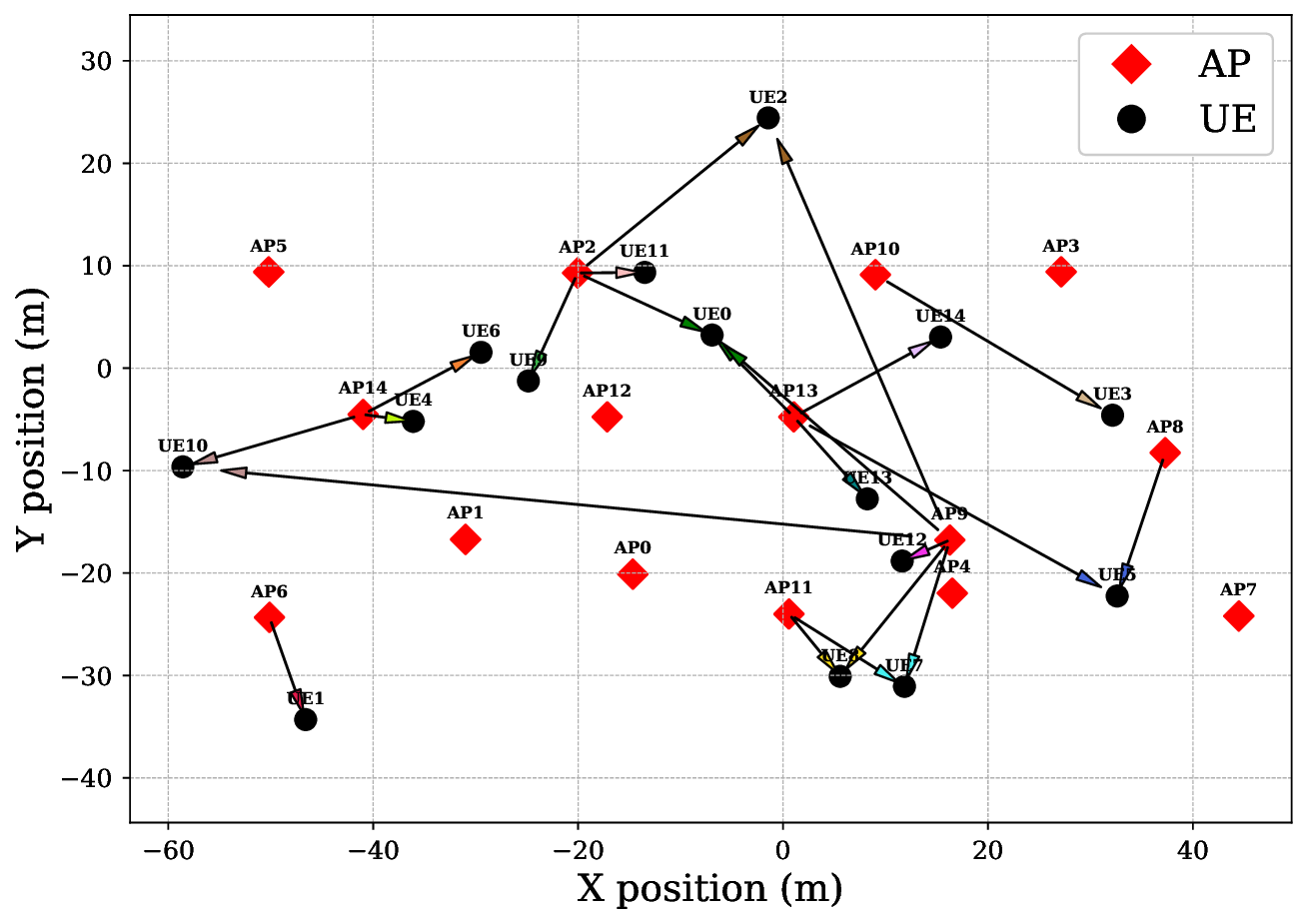}
             \caption{}
            \label{fig:Rice_clusters}
    \end{subfigure}
    \caption{(a) evaluates the power saving capability and energy efficiency gain via EA-based solution with spatially correlated Rician channels under different loading scenarios while (b) shows an example of cluster formation via EA-based algorithm with $K=15$ at a given scenario.}
    \label{fig:Rice_P_EE_effective_clusters}
\end{figure}
\subsection{Discussion}
The EA-based algorithm optimizes network performance by selectively engaging only the necessary APs, thereby reducing energy consumption. Instead of engaging all available APs, the algorithm evaluates the QoS satisfaction for each UE and the energy efficiency of its cluster by identifying and establishing the most social favorable association pairs during each iteration. In Fig. \ref{fig:Rice_P_EE_vs_effective_P_EE_EA}, 
we evaluate the power saving capability and energy efficiency gain when non-engaged \acp{AP} are turned off. In particular, we implement a symbol shutdown resulting in a $30\%$ reduction of the baseline power $P^{\rm{AAU, fix}}$ at each unassociated AP \cite{Machine_Learning_and_Analytical_Power_Consumption_Models_for_5G_Base_Stations}. 
In this case, we denote by $P_{\rm eff}$ and $\mathrm{EE}_{\rm eff}$ the effective power consumption and effective network energy efficiency, respectively. We show that as the number of \acp{UE} increases, additional \acp{AP} are engaged to meet the growing demand with the presence of non-engaged APs. By deactivating unassociated APs, the network's energy efficiency is enhanced, as unnecessary energy expenditure is avoided. However, maintaining \acp{AP} active to serve only a single UE can result in inefficient energy use (e.g. \acp{AP} $6$, $8$ and $10$ in Fig. \ref{fig:Rice_clusters}). This highlights the importance of implementing sleep mode strategies \cite{Sleep_Mode_Strategies_for_Energy_Efficient_Cell_Free_Massive_MIMO_in_5G_Deployments} to optimize energy consumption while still meeting the QoS requirements of UEs. Additionally, the proposed solution must adapt in real-time to network dynamics including channel variations and user mobility. By considering the QoS requirements of UEs, the algorithm can dynamically select the optimal set of \acp{AP} to keep active, balancing the trade-off between QoS satisfaction and energy efficiency.

\begin{figure}[!t]
    \centering
       \begin{subfigure}[t]{0.49\columnwidth}
        \centering
             \includegraphics[width=\columnwidth]{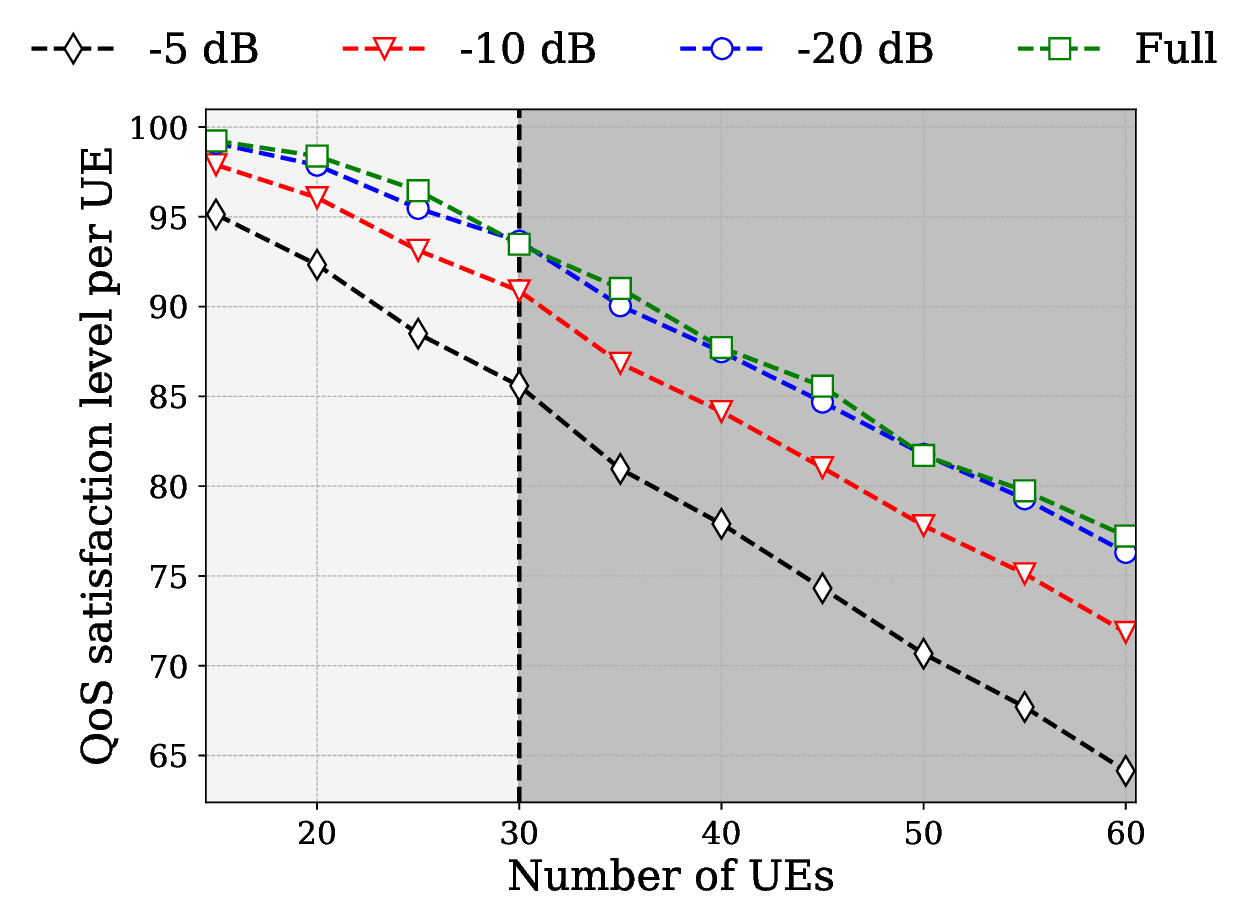}
             \caption{}
            \label{fig:Rice_different_SNR_Kappa}
    \end{subfigure}
    \hfill
    \begin{subfigure}[t]{0.49\columnwidth}
        \centering
             \includegraphics[width=\columnwidth]{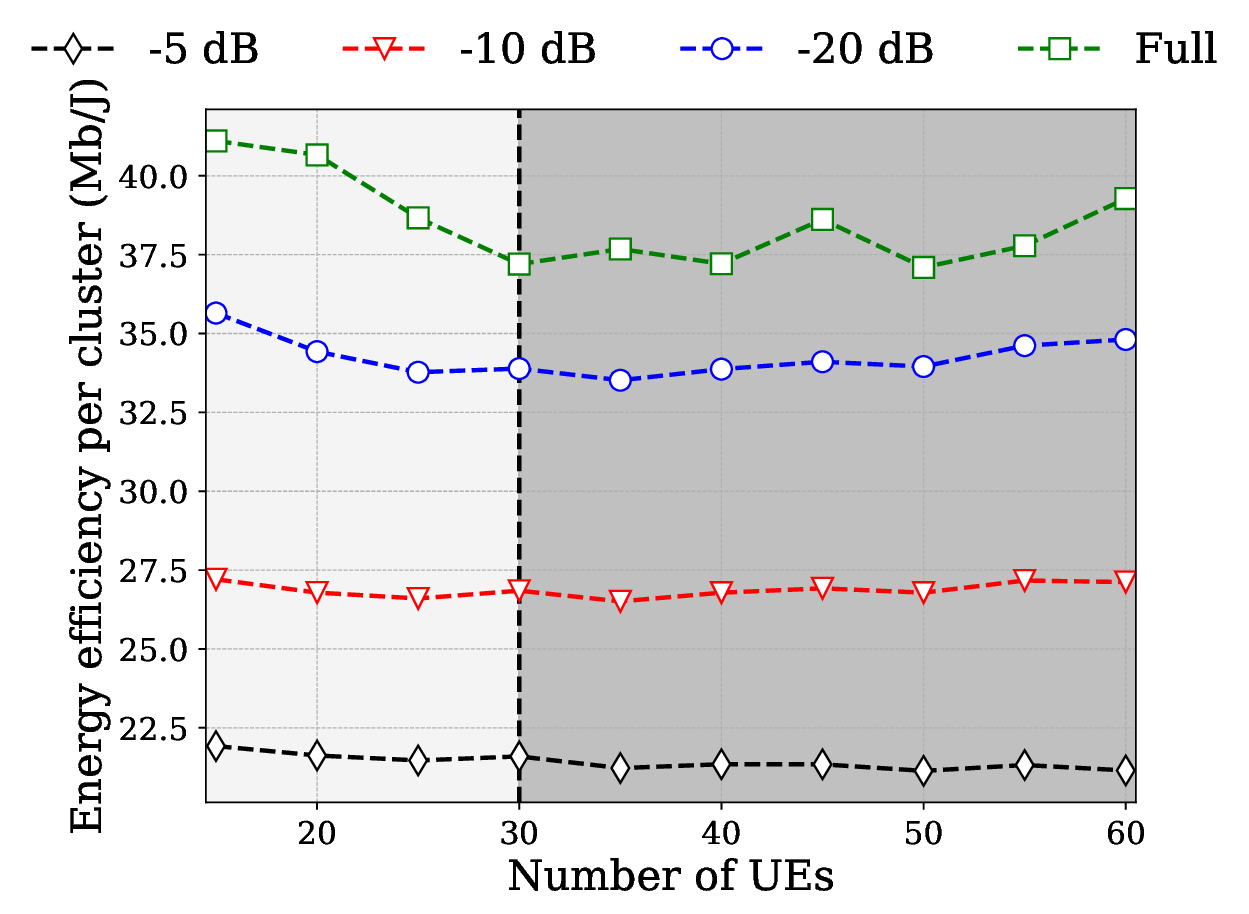}
             \caption{}
            \label{fig:Rice_different_SNR_EE_cluster}
    \end{subfigure}
    \caption{Impact of channel estimation accuracy on QoS satisfaction and energy efficiency per \ac{UE} with spatially correlated Rician channels under different loading scenarios.}
    \label{fig:Rice_different_SNR}
\end{figure}

In realistic wireless networks, each AP performs local channel estimation to determine its link quality with UEs in the network. As a result, both the clustering process and the precoding scheme are potentially highly sensitive to the quality of the channel estimation. Specifically, preference lists derived from erroneous channel estimates may no longer reflect the true preferences of \acp{AP} and \acp{UE}, leading to poor connections. Furthermore, the performance of local ZF to mitigate interference between UEs served by the same AP may be significantly degraded when perfect CSI is not available at the AP. Fig. \ref{fig:Rice_different_SNR} evaluates the performance of our EA-based scheme under different values of \ac{NMSE} in the \ac{CSI}. The figure highlights the significant impact of estimation quality on both QoS satisfaction and energy efficiency per cluster. In particular, at an \ac{NMSE} of $-20$ dB, our proposed EA-based clustering provides performance that are comparable with the ideal case with full knowledge of CSI, indicating that high accurate channel estimation can preserve most of the clustering and ZF precoding benefits. In contrast, higher \ac{NMSE} levels introduce significant channel estimation errors, which affect cluster formation quality, thus leading to performance degradation.

\section{Conclusion}\label{sec:conclusion}
In this paper, we studied the user association problem in \ac{CF-MIMO} networks to jointly maximize the QoS satisfaction per UE and its local energy efficiency, while considering the \textit{sociality regime} of \acp{UE} and their clusters, being either selfish, egalitarian, or altruistic. To find a balance between these two conflicting interests, we formulate this problem as a many-to-many \textit{social} matching game and propose two algorithms based on \ac{DA} and \ac{EA}. Numerical results highlight the impact of the sociality regime on performance in terms of perceived QoS and local and global energy efficiency. In particular, adopting egalitarian regime by \acp{UE} and their clusters of \acp{AP} emerges as the most advantageous combination in terms of QoS satisfaction per UE and local energy efficiency, due to its fairness and equitable resource distribution among UEs. Under this specific sociality regime combination, the EA-based scheme demonstrates superior performance, providing the best perceived QoS and energy efficiency per UE compared to the DA-based solution in dense networks, while also requiring less computational resources. Moreover, the EA-based approach considerably reduces the number of AP-UE connections by at least $44\%$ compared to DA-based algorithm, thereby significantly limiting the fronthaul signaling and processing delay. 

In future work, we will include sleep mode mechanisms to further improve the energy efficiency of the network. Moreover, we aim to propose an effective handover algorithm that adapts in real-time to user mobility and channel variations.


\appendices
\section{Proof of Proposition \ref{prop:ue_social_factor}}\label{apx:ue_social_factor}
We have:
\begin{equation}\label{eq:sum_UEs_utilities}
    \begin{aligned}
        \displaystyle\sum_{k \in \mathcal{K}} \gamma_{k}^{\alpha}(\param) &= \displaystyle\sum_{k \in \mathcal{K}} \left[ \alpha\rho_{k}(\param) + \frac{1 - \alpha}{K - 1}\sum_{\substack{k' \in \mathcal{K}\backslash k}} \rho_{k'}(\param) \right] \\
        &= \alpha \displaystyle\sum_{k \in \mathcal{K}} \rho_{k}(\param) +  \frac{1 - \alpha}{K - 1} \displaystyle\sum_{k \in \mathcal{K}} \sum_{\substack{k' \in \mathcal{K}\backslash k}} \rho_{k'}(\param)
    \end{aligned}
\end{equation}

or we know that $\displaystyle\sum_{\substack{k' \in \mathcal{K}\backslash k}} \rho_{k'}(\param) = \Gamma(\param) - \rho_{k}(\param)$. 

Thus, $ \displaystyle\sum_{k \in \mathcal{K}} \sum_{\substack{k' \in \mathcal{K}\backslash k}} \rho_{k'}(\param) = (K-1) \Gamma(\param)$. 

Then, Eq. \eqref{eq:sum_UEs_utilities} becomes:
\begin{equation}
    \begin{aligned}
        \displaystyle\sum_{k \in \mathcal{K}} \gamma_{k}^{\alpha}(\param) &= \alpha \Gamma(\param) + \frac{1 - \alpha}{K - 1} (K-1) \Gamma(\param) \\
        &= \Gamma(\param)
    \end{aligned}
\end{equation}

\section{Proof of Proposition \ref{prop:ap_social_factor}}\label{apx:ap_social_factor}
We have:
\begin{equation}\label{eq:sum_APs_CPU_utilities}
    \begin{aligned}
       \displaystyle\sum_{\substack{k \in \mathcal{K}}} \mathcal{E}_{k}^{\beta}(\param) =  \sum_{\substack{k \in \mathcal{K}}}\left[\beta \mathrm{EE}_{k}(\param) + \frac{1 - \beta}{K - 1}\sum_{\substack{k' \in \mathcal{K}\backslash k}} \mathrm{EE}_{k'}(\param)\right] \\
       = \beta \sum_{\substack{k \in \mathcal{K}}} \mathrm{EE}_{k}(\param) + \frac{1 - \beta}{K - 1} \sum_{\substack{k \in \mathcal{K}}} \displaystyle\sum_{\substack{k' \in \mathcal{K}\backslash k}} \mathrm{EE}_{k'}(\param)
    \end{aligned}
\end{equation}

or we know that $\displaystyle\sum_{\substack{k' \in \mathcal{K}\backslash k}} \mathrm{EE}_{k'}(\param) = \mathrm{EE}(\param) - \mathrm{EE}_{k}(\param)$. 

Thus, $\displaystyle\sum_{\substack{k \in \mathcal{K}}} \displaystyle\sum_{\substack{k' \in \mathcal{K}\backslash k}} \mathrm{EE}_{k'}(\param) = (K - 1) \mathrm{EE}(\param)$. 

Then, Eq. \eqref{eq:sum_APs_CPU_utilities} becomes:
\begin{equation}
    \begin{aligned}
        \displaystyle\sum_{\substack{k \in \mathcal{K}}} \mathcal{E}_{k}^{\beta}(\param) &= \beta \mathrm{EE}(\param) + \frac{1 - \beta}{K - 1} (K - 1) \mathrm{EE}(\param) \\
        &= \mathrm{EE}(\param)
    \end{aligned}
\end{equation}

\section{Cluster power consumption}
\label{apx:clusterPower}
This Appendix details the computation of the power $P_{\mathcal{C}_{k}^{\rm UE}}(\param)$ consumed by the cluster of \acp{AP} $\mathcal{C}_{k}^{\rm UE}$ cooperating to serve UE $k\in\mathcal{A}^{+}$. It comprises the power consumed by the CPU and the \acp{AP} serving the \ac{UE}. 
\begin{equation}\label{eq:cluster_power}
    P_{\mathcal{C}_{k}^{\rm UE}}(\param) = P_{k}^{\rm CPU}(\param)+\sum_{m \in \mathcal{M}}c_{k,m}P_{k,m}(\param).
\end{equation}

In Eq. \eqref{eq:cluster_power}, $P_{k}^{\rm CPU}(\param)$ is the power consumed by the CPU to serve \ac{UE} $k$, which we compute as: 
\begin{equation}\label{eq:cpu_cluster_power}
    P_{k}^{\rm CPU}(\param) = P_{\rm UE}^{\rm CPU, fix}(\param) + R_{k}(\param)P^{\rm CPU,enc}, ~\forall k \in \mathcal{A}^{+}.
\end{equation}
Also, $P_{k,m}(\param)$ denotes the power consumption of \ac{AP} $m$ serving \ac{UE} $k$; $\forall m \in \mathcal{M}^{+}$, its expression reads as:
\begin{equation}\label{eq:ap_cluster_power}
    P_{k,m}(\param) = P_{m}^{\rm UE, fix}(\param) + \frac{1}{\zeta}P_{k,m}^{\rm tx}(\param) + P^{\rm FH,prec}.
\end{equation}

In Eq. \eqref{eq:cpu_cluster_power}, the aggregate fixed power consumption of the CPU to serve \ac{UE} $k$ $P_{\rm UE}^{\rm CPU, fix}$ is equally distributed among the total number of \acp{UE} connected in the network:
\begin{equation}\label{eq:CPU_fix_power_UE}
    P_{\rm UE}^{\rm CPU, fix}(\param) = \frac{P^{\rm CPU,fix}}{|\mathcal{A}^{+}|}.
\end{equation}
Similarly to Eq. \eqref{eq:ap_cluster_power}, the fixed power consumption of \ac{AP} $m$ to serve \ac{UE} $k$ $P_{m}^{\rm UE, fix}$ is determined by equally distributing the total fixed power consumption by that \ac{AP} among all the \acp{UE} it serves:
\begin{equation}\label{eq:AP_fix_power_UE}
   P_{m}^{\rm UE, fix}(\param) = \frac{P^{\rm AAU,fix} + P^{\rm FH,fix}}{|\mathcal{C}_{m}^{\rm AP}|}.
\end{equation}

\bibliographystyle{ieeetr}
\bibliography{biblio}

\end{document}